\documentclass[%
 preprintnumbers,
 amsmath,amssymb,
 prl, 
 twocolumn]{revtex4-2}

\usepackage{SIunits}
\usepackage{graphicx}
\usepackage{xspace}
\usepackage{braket}
\usepackage{bm}
\usepackage{placeins}
\usepackage{leftidx}

\usepackage[utf8]{inputenc}

\usepackage[usenames,dvipsnames]{color}

\usepackage[pdftex, colorlinks=true, urlcolor=blue, linkcolor=blue, citecolor=blue, pdfstartview=FitH, plainpages=false]{hyperref}

\newcommand{\kb}{\mathbf{k}}
\newcommand{\qb}{\mathbf{q}}
\newcommand{\qq}{\mathbf{q}}

\newcommand{\rb}{\mathbf{r}}
\newcommand{\Gb}{\mathbf{G}}

\newcommand{\enk}{\varepsilon_{n\kb}}
\newcommand{\emkq}{\varepsilon_{m\kb+\qb}}

\newcommand{\wnuq}{\omega_{\qb\nu}}
\newcommand{\qnu}{{{\qb\nu}}}

\newcommand{\vnka}{\mathrm{v}_{n\kb,\alpha}}
\newcommand{\vnkb}{\mathrm{v}_{n\kb,\beta}}

\newcommand{\Lc}{\mathcal{L}}
\newcommand{\Sc}{\mathcal{S}}
\newcommand{\abinit}{\textsc{Abinit}\xspace}

\newcommand{\gkkp}{g_{mn\nu}(\kb,\qb)}

\newcommand{\abinitio}{\textit{ab initio}\xspace}
\newcommand{\grid}[1]{${#1}\times{#1}\times{#1}$}

\newcommand{\rr}{{\mathbf r}}

\newcommand{\GG}{{\bf G}}

\newcommand{\KS}{{\text{KS}}}

\newcommand{\Zstar}{{\bf Z}^*}
\newcommand{\BZ}{{\text{BZ}}}
\newcommand{\epsinf}{{\bm{\epsilon}}^\infty}
\newcommand{\zero}{\mathbf{0}}

\newcommand{\dd}{{\,\text{d}}}

\begin{document}

\title{
\rlap{Electron-Phonon Beyond Fr\"ohlich:
Dynamical Quadrupoles in Polar and Covalent Solids}
}

\date{\today}
\author{Guillaume Brunin$^1$}
\author{Henrique Pereira Coutada Miranda$^1$}
\author{Matteo Giantomassi$^1$}
\author{Miquel Royo$^2$}
\author{Massimiliano Stengel$^{2,3}$}
\author{Matthieu J. Verstraete$^{4,5}$}
\author{Xavier Gonze$^{1,6}$}
\author{Gian-Marco Rignanese$^1$}
\author{Geoffroy Hautier$^1$}
\email{Corresponding author: geoffroy.hautier@uclouvain.be}

\affiliation{$^1$UCLouvain, Institute of Condensed Matter and Nanosciences (IMCN), Chemin des \'Etoiles~8, B-1348 Louvain-la-Neuve, Belgium}
\affiliation{$^2$Institut de Ci\`encia de Materials de Barcelona (ICMAB-CSIC), Campus UAB, 08193 Bellaterra, Spain}
\affiliation{$^3$ICREA-Instituci\'o Catalana de Recerca i Estudis Avan\c{c}ats, 08010 Barcelona, Spain}
\affiliation{$^4$NanoMat/Q-Mat/CESAM, Universit{\'e} de Li{\`e}ge (B5), B-4000 Li{\`e}ge, Belgium}
\affiliation{$^5$Catalan Institute of Nanoscience and Nanotechnology (ICN2), Campus UAB, 08193 Bellaterra, Spain}
\affiliation{$^6$Skolkovo Institute of Science and Technology, Moscow, Russia}

\begin{abstract}

We include the treatment of quadrupolar fields beyond
the Fr\"ohlich interaction in the first-principles electron-phonon vertex in semiconductors.
Such quadrupolar fields induce long-range interactions that have to be taken into account for accurate physical results.
We apply our formalism to Si (nonpolar), GaAs, and GaP (polar) and demonstrate that electron mobilities show large errors if dynamical quadrupoles are not properly treated.

\end{abstract}

\maketitle


The electron-phonon (e-ph) interaction plays a key role in the description of various physical phenomena (e.g., electronic transport, phonon-assisted light absorption, phonon-mediated superconductivity)~\cite{Giustino2017}.
In state-of-the-art \abinitio methods, the e-ph coupling is described 
within density functional theory (DFT) by expanding the Kohn-Sham (KS) effective potential~\cite{Kohn1965} 
in the nuclear displacements, 
while vibrational properties are obtained using density-functional perturbation theory
(DFPT)~\cite{Gonze1997,Baroni2001}.
This DFPT-based computational scheme 
enables the calculation of screened \textit{e}-ph 
matrix elements at a microscopic level
with \abinitio quality~\cite{Giustino2017}.
However, accurate \textit{e}-ph related properties
require a description of the coupling 
on very dense reciprocal-space wave vector grids across the full Brillouin zone (BZ)  
thus making direct \abinitio \textit{e}-ph computations in real materials costly and sometimes even impracticable.

Despite formal differences in the treatment of the electron wave function, 
all approaches proposed so far
rely on the spatial localization of the 
\textit{e}-ph coupling
to interpolate it with respect to the phonon wave vector $\qq$~\cite{Giustino2007,Sjakste2015,Ponce2016,Agapito2018}.
The fundamental physical assumption is that a relatively small real-space supercell can capture the full strength of the coupling. For metals this is trivial as screening lengths are very short.
However, in semiconductors and insulators, the incomplete screening of the potential generated by the atomic displacements
leads to long-range (LR) interactions.
For polar materials, these interactions show up in the long-wavelength limit ($\qq \rightarrow \zero$) for instance in the LO-TO splitting of the optical frequencies~\cite{Born1954}
and the Fr\"ohlich divergence of the \textit{e}-ph matrix elements~\cite{Frohlich1954}.
The analytical treatment of the LR dipole-dipole interaction 
to obtain a reliable long-wavelength dynamical matrix
was developed in the early days of DFPT~\cite{Gonze1997,Baroni2001}.
More recently the Fr\"ohlich contribution to the LR \textit{e}-ph matrix elements was generalized
to anisotropic materials~\cite{Verdi2015,Sjakste2015}, opening up a first avenue for computations of \textit{e}-ph effects in polar materials~\cite{Giustino2017}.

In this Letter, we go beyond these seminal contributions, demonstrate the importance of 
the next-to-leading order terms
derived by Vogl~\cite{Vogl1976}, and explain their physical origin. In this new approach, the LR contributions are expressed in terms
of the high-frequency dielectric tensor, Born effective charges (dipole potential),
dynamical quadrupoles~\cite{Royo2019,Stengel2013}
(quadrupole potential)
and the response to a homogeneous static electric field (local-field potential, also quadrupolar)~\cite{Gonze1997,Baroni2001}. 
Despite their importance in describing certain piezoelectric properties~\cite{Royo2019,Martin1972}, 
quadrupolar interactions have never been considered in first-principles \textit{e}-ph calculations. In nonpolar materials, dynamical quadrupoles represent the leading contribution to the LR potential. Even in polar materials, where the focus has been on the divergent Fr\"ohlich interaction for LO modes~\cite{Verdi2015}, the quadrupolar field dominates the LR potential for the TO and acoustic modes~\cite{Vogl1976}.

We first describe the physical origin and LR nature of the quadrupolar interaction. We then demonstrate that substantial errors in carrier mobility in well-known polar and nonpolar semiconductors are introduced if the quadrupolar interaction is not properly accounted for. We show results for the classical semiconductors Si, GaAs, and GaP. As quadrupolar interactions have not been considered so far, our analysis
indicates that many previous \abinitio \textit{e}-ph calculations for semiconductors
should be critically reconsidered.


The key ingredients in \textit{e}-ph computations are the coupling matrix elements 
$\gkkp = \braket{\psi_{m\kb+\qb}|\Delta_\qnu V|\psi_{n\kb}}$ 
with $\psi_{n\kb}$ the $n\kb$ Bloch state and $\Delta_\qnu V$ the first-order variation of the KS potential $V$
due to a phonon mode of wave vector $\qq$ and branch index $\nu$~\cite{Giustino2017}.
The scattering potential is defined as
\begin{equation}
    \Delta_\qnu V = 
    \dfrac{1}{\sqrt{2\wnuq}} \sum_{\kappa\alpha} 
    \dfrac{{e}_{\kappa\alpha,\nu}(\qb)}{\sqrt{M_\kappa}} V_{\kappa\alpha,\qb}(\rb)
\end{equation}
with $\wnuq$ the phonon frequency and ${e}_{\kappa\alpha,\nu}(\qb)$ the $\alpha$-th Cartesian component of the phonon eigenvector 
for atom $\kappa$ of mass $M_\kappa$ in the unit cell.
$V_{\kappa\alpha,\qb}(\rb)$ is the potential obtained from DFPT~\footnote{More specifically, $V_{\kappa\alpha,\qb}(\rb) \equiv e^{i \qq \cdot \rr} \partial_{\kappa\alpha,\qq} v^\KS(\rb)$, with $\partial_{\kappa\alpha,\qq} v^\KS(\rb)$ the lattice-periodic part computed with DFPT~\cite{Brunin2019prb}}.
Following the approach adopted in the literature~\cite{Verdi2015,Sjakste2015},
$V_{\kappa\alpha,\qb}$
is separated into short-range (SR, $\Sc$) and LR ($\Lc$) contributions: 
$ V_{\kappa\alpha,\qb}(\rb) = V^{\Sc}_{\kappa\alpha,\qb}(\rb) + V^{\Lc}_{\kappa\alpha,\qb}(\rb)$.
The latter, $V^{\Lc}$, 
is supposed to include all the LR components so that $V^{\Sc}$
is smooth in $\qq$ space and therefore tractable with Fourier interpolation. In the interpolation algorithm, $V^{\Lc}$ is first subtracted from the DFPT potentials, 
then the Fourier interpolation is performed on the SR part only.
$V^{\Lc}$ evaluated at the arbitrary $\qq$ point is finally added back~\cite{Brunin2019prb}.
In polar materials, the leading contribution to the LR part stems from the diverging Fr\"ohlich-like potential~\cite{Verdi2015,Sjakste2015}:
\begin{equation}
    V^{\Lc(F)}_{\kappa\alpha,\qb}(\rb) = 
    \frac{4\pi}{\Omega} \sum_{\Gb\neq\mathbf{-q}}
    \frac{i (q_{\beta}+G_{\beta}) 
    Z^*_{\kappa\alpha,\beta} e^{i (q_\eta + G_\eta) (r_\eta - \tau_{\kappa\eta})}}
    {(q_{\delta}+G_{\delta})\epsilon^{\infty}_{\delta\delta'}(q_{\delta'}+G_{\delta'})}
    \label{eq:LRFpot}
\end{equation}
with $\Omega$ the unit cell volume, $\GG$ the reciprocal lattice vectors, $\Zstar_{\kappa}$ the Born effective charge tensor, $\epsinf$ the high-frequency dielectric tensor, and
$\bm{\tau}_{\kappa}$ the position of the $\kappa$-th atom in the unit cell.
The summation over repeated indices ($\beta$, $\eta$, $\delta$, and $\delta^\prime$) is implied in Eq.~\eqref{eq:LRFpot} and in the following, unless the sum is explicitly written.

\begin{figure}
    \centering
    \includegraphics[clip,trim=0cm 0cm 0.8cm 0cm,width=0.48\textwidth]{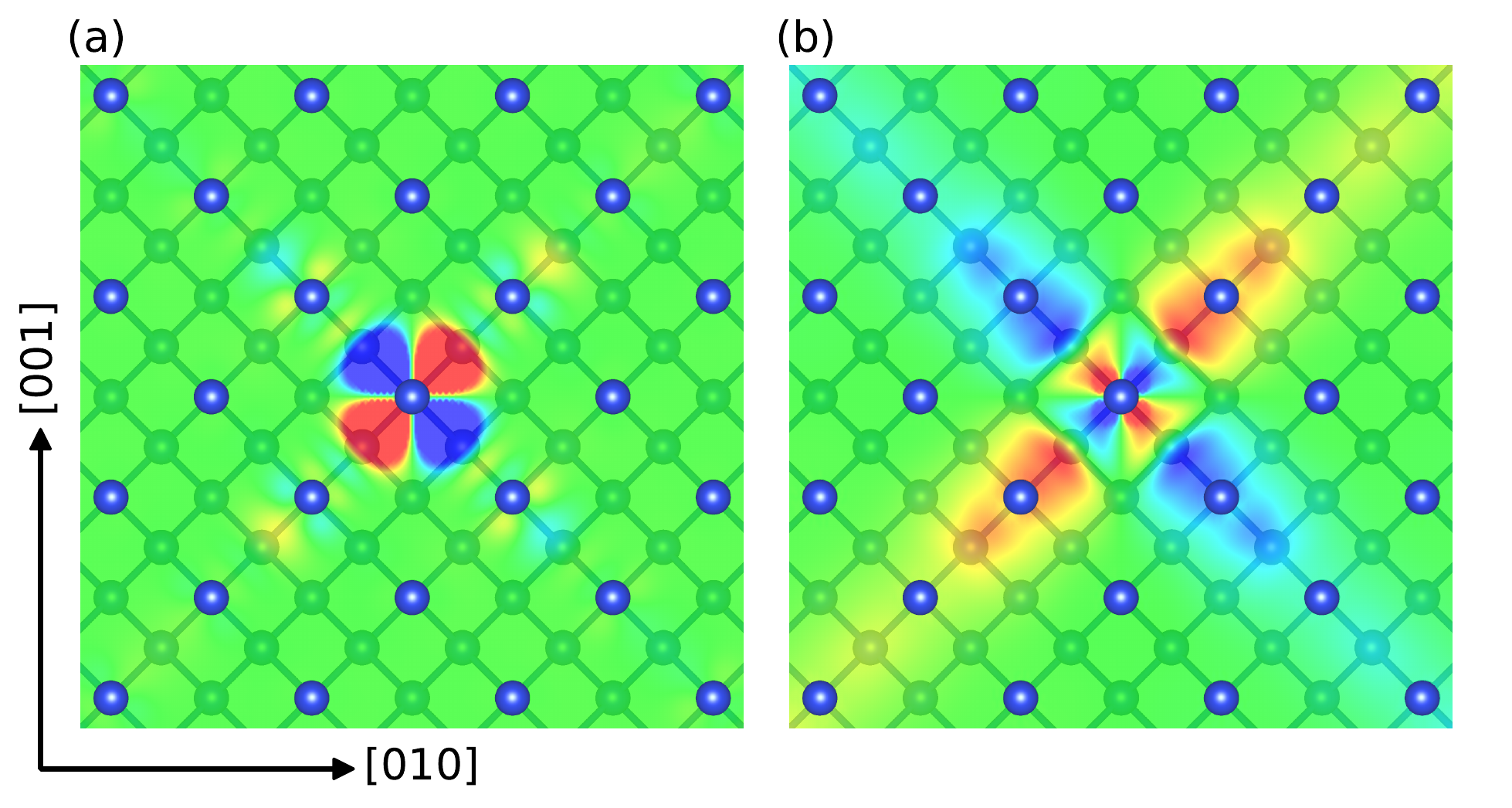}
    \caption{(a) Electronic charge density variation induced by a single
    atomic displacement in Si, 
    along the [100] direction of the conventional cell ({\it perpendicular} to the represented plane).
    Negative (positive) variations are shown in red (blue), while green regions correspond to absolute values smaller
    than 10\% of the in-plane maximum.
    (b) Scattering potential
    induced by the atomic displacement with the same color map as in (a).
    SR dipolelike contributions are absent in this plane~\cite{supplemental_prl}.
    }
\label{fig:quadrupole}
\end{figure}

Most investigations so far have focused on the treatment of Eq.~\eqref{eq:LRFpot}. 
However, as discussed by Vogl~\cite{Vogl1976} and derived in a DFPT context in our accompanying paper~\cite{Brunin2019prb}, a careful analysis of the asymptotic behavior of the scattering potential
in the long-wavelength limit reveals the presence of additional 
LR terms besides Eq.~\eqref{eq:LRFpot}.
To the contrary of the $1/q$ nature of Eq.~\eqref{eq:LRFpot}, these 
additional terms 
are finite for $\qq \rightarrow \zero$ but their nonanalytic behavior (angular discontinuities) 
yields LR 
scattering potentials and associated \textit{e}-ph matrix elements even when the dipole interaction given by Eq.~\eqref{eq:LRFpot} 
is properly accounted for.
Both dipole and quadrupole terms can be included in the LR potential using the generalized expression~\cite{Brunin2019prb}:
\begin{widetext}
    \begin{equation}
        V^{\Lc}_{\kappa\alpha,\qb}(\rb) = 
        \frac{4\pi}{\Omega} \sum_{\Gb\neq\mathbf{-q}}
        \frac{ i(q_{\beta}+G_{\beta}) 
        Z^*_{\kappa\alpha,\beta} 
        +
        (q_{\beta}+G_{\beta})(q_{\gamma}+G_{\gamma})
        (Z^*_{\kappa\alpha,\beta}
        v^{\text{Hxc},\cal{E}_{\gamma}}(\rr) +
        \frac{1}{2}Q_{\kappa\alpha}^{\beta\gamma})
        }
        {
        {(q_{\delta}+G_{\delta})\epsilon^{\infty}_{\delta\delta'}(q_{\delta'}+G_{\delta'})}}
        e^{i (q_\eta + G_\eta) (r_\eta - \tau_{\kappa\eta})},
    \label{eq:LRpot}
    \end{equation}
\end{widetext}
where $\mathbf{Q}_{\kappa\alpha}$ is the dynamical quadrupole tensor and $v^{\text{Hxc},\cal{E}}$ is the change of the Hartree and exchange-correlation potential with respect to the electric field ${\cal{E}}$ in Cartesian coordinates. 
In Eq.~\eqref{eq:LRpot}, the term related to $\mathbf{Q}_{\kappa\alpha}$ 
is nonzero even in nonpolar semiconductors while the contribution associated with 
$\cal{E}$ is present only in systems with nonzero Born effective charges. 
A quadrupole can indeed appear even in nonpolar situations: when mirror and/or inversion symmetries are broken at a given atomic site, the charge induced by atomic motion can acquire an inversion-even and hence quadrupolar contribution. This quadrupolar effect is even present in elemental crystals with vanishing dipoles.  
This effect is illustrated in Fig.~\ref{fig:quadrupole}(a) reporting the variation of the charge density induced by a single atomic displacement in a 250-atoms supercell of Si. 
Such \emph{dynamical} quadrupoles should not be confused with the static quadrupolar moment that characterizes certain molecules, e.g., CO$_2$~\cite{Graham1989}.
Regardless of its static or dynamic nature, 
a localized charge quadrupole results in a 
LR scattering potential that
extends over many unit cells of the unperturbed
crystal, see Fig.~\ref{fig:quadrupole}(b).
In the region around the displaced atom,
this term is relatively small compared to the SR part
but it becomes the dominant 
contribution at large distances.
Its anisotropic and macroscopic character in real space determines the $\qq \rightarrow \zero$ behavior of the \textit{e}-ph matrix elements in nonpolar materials,
and constitutes an important contribution in polar materials.
By contrast, the $v^{\text{Hxc},\cal{E}}$ contribution is a secondary, indirect effect of the Fr\"ohlich interaction: a local-field potential modification due to the 
charge density change induced by 
the Fr\"ohlich dipolar electric field. 
Thus, it vanishes when $\Zstar_{\kappa}$ vanishes.

There is no \textit{a priori} rigorous justification for ignoring 
LR anisotropic contributions to the \textit{e}-ph scattering generated by 
the quadrupolar terms. This is especially critical because \textit{e}-ph properties are rather sensitive to the 
behavior of the 
potential in the macroscopic regime. 
We first 
gauge the 
importance of these effects 
on the \textit{e}-ph scattering potentials. 
The effects on electron mobilities are discussed afterwards. 
Our study is based on a new \textit{e}-ph implementation in \abinit detailed in 
our accompanying paper~\cite{Brunin2019prb}. 
The numerical values of $\mathbf{Q}_{\kappa\alpha}$ are computed using the recent implementation of Royo \textit{et al.}~\cite{Royo2019}, 
now integrated with the e-ph part of \abinit~\cite{Gonze2016,Gonze2019,Romero2020,Brunin2019prb}. 
Following previous works~\cite{Eiguren2008,Chaput2019}, 
we treat the electron wave functions exactly and employ a Fourier-transform-based scheme for the scattering potentials~\cite{Eiguren2008}. 
This avoids a transformation to localized orbitals, and has
the important advantage of being systematic and automatic. 

\begin{figure}
    \centering
    \includegraphics[clip,trim=0.2cm 0.2cm 0.2cm 0.2cm,width=.35\textwidth]{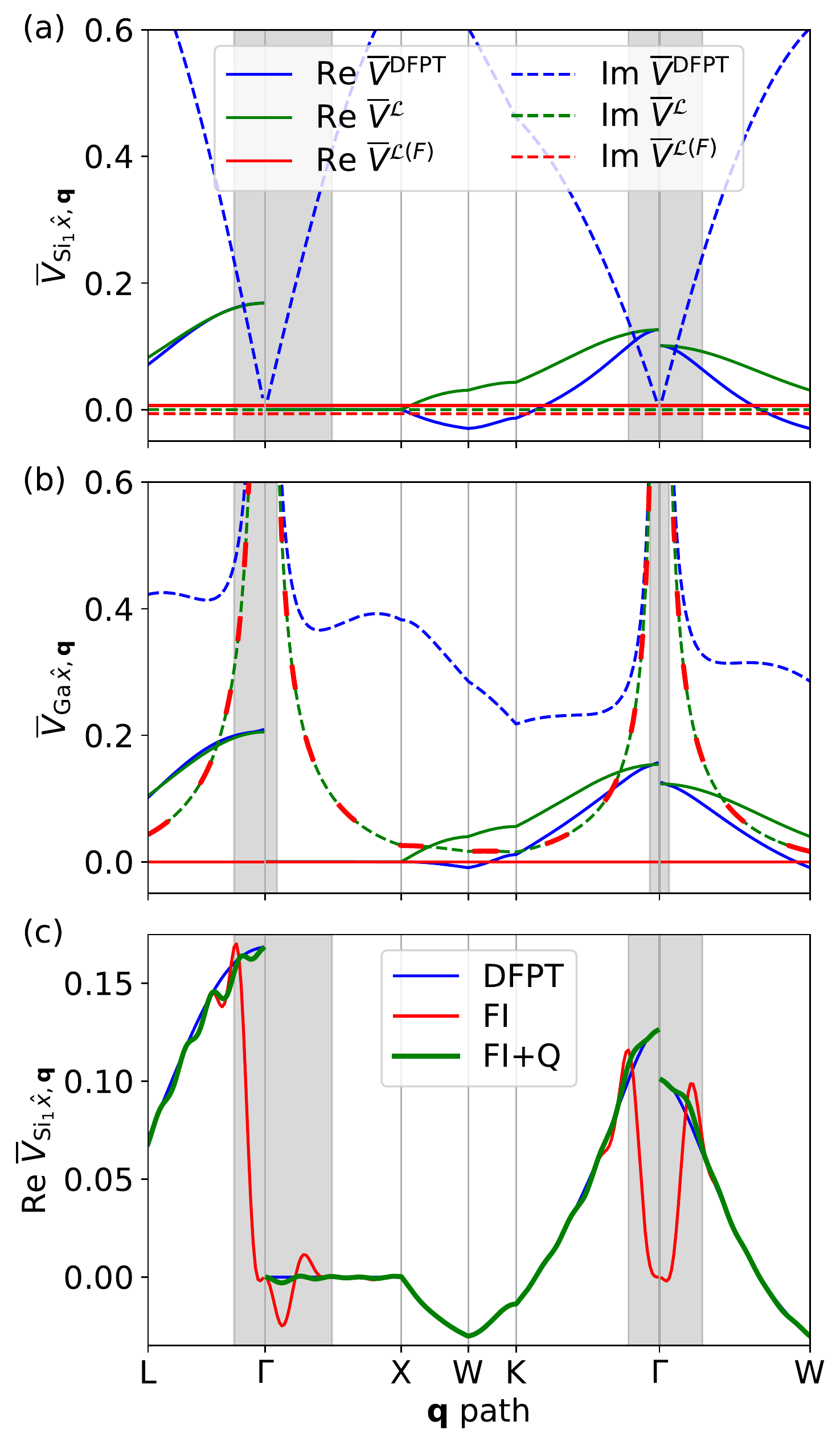}
    \caption{Comparison of the unit-cell averaged exact DFPT potentials ($\overline{V}^\text{DFPT}$, blue lines) with the models of Eqs.~\eqref{eq:LRFpot} ($\overline{V}^{\Lc(F)}$, red lines) and ~\eqref{eq:LRpot} ($\overline{V}^{\Lc}$, green lines)  in (a) Si and (b) GaAs.
    We consider the first reduced component $\hat{x}$ of the DFPT potentials for (a) the Si 
    and (b) the Ga atoms located at (0,0,0).
    In (a), the red lines have been slightly shifted upwards and downwards for readability.
    (c) Fourier interpolation 
    of the potentials shown in (a) with (FI+Q) or without (FI) the treatment of the quadrupolar interaction, as explained in the text.
    The potentials are given in Hartree.
    The path has been sampled with 278 $\qq$ points.
    The gray areas around $\Gamma$ represent the regions relevant for intravalley scattering, as dictated by Eq.~\eqref{eq:tau}.
    }
    \label{fig:potentials}
\end{figure}

In Figs.~\ref{fig:potentials}(a) and \ref{fig:potentials}(b) we plot the unit-cell averaged lattice-periodic part of the scattering potential,
\begin{equation}
    \bar{V}_{\kappa\alpha,\qb} =
    \dfrac{1}{\Omega}\int_\Omega \dd\rr\,V_{\kappa\alpha,\qb}(\rb) e^{-i\qq\cdot\rr},
\end{equation}
for selected atomic perturbations in Si and GaAs, along a high-symmetry $\qb$ path. 
The \emph{exact} DFPT results (blue lines) are compared with those obtained with the models with (green) or without (red) quadrupole corrections in Eq.~\eqref{eq:LRpot}~\cite{supplemental_prl}. 
In Si, the Born effective charges are zero and the imaginary part of the potential does not diverge for $\qq \rightarrow \zero$ 
(see dashed lines in Fig.~\ref{fig:potentials}(a)). 
In GaAs, the Fr\"ohlich-like model in Eq.~\eqref{eq:LRFpot} correctly describes the divergence 
of the imaginary part of the potential close to $\Gamma$ 
(see red dashed line in Fig.~\ref{fig:potentials}(b)). 
In both materials, however, the real part of the potential (solid lines) presents discontinuities for $\qq \rightarrow \zero$. 
Note that 
the Fr\"ohlich term alone completely misses this nonanalytic behavior. 
On the contrary,  
if the quadrupolar 
contributions are included through Eq.~\eqref{eq:LRpot}, 
the LR model reproduces these discontinuities 
as shown by the solid green lines in Figs.~\ref{fig:potentials}(a) and \ref{fig:potentials}(b). 
Figure~\ref{fig:potentials}(c) shows the real part of the average of the scattering potential in Si interpolated from a \grid{9} 
$\qb$-point grid onto the same $\qb$ path as in Fig.~\ref{fig:potentials}(a). 
If the 
LR quadrupolar terms are not removed from the input DFPT potentials, Fourier aliasing introduces unphysical sharp oscillations for small $\qq$ (see red line, FI). 
The correct behavior is 
obtained 
only when 
quadrupolar contributions are 
properly treated 
(see green line, FI+Q). 
In the latter case, 
small oscillations 
between the \abinitio $\qq$ points 
are still visible 
when a \grid{9} $\qq$ mesh is used. 
However, convergence studies~\cite{supplemental_prl} reveal that these wiggles have a limited effect on the final electron mobility (less than 0.5\% difference between the 
mobility obtained for a \grid{9} and an \grid{18} \abinitio $\qq$ mesh) 
and they can be converged away. 

It is worth stressing that these considerations hold for any approach employing Fourier-based interpolations. 
The discontinuity of the matrix elements at $\Gamma$ 
and the discrepancy between the interpolant and the \emph{exact} DFPT results
in the region around $\Gamma$
have already been noticed for Si and diamond~\cite{Agapito2018}. 
The resulting error was considered harmless under the assumption that it is always possible to improve the accuracy 
of the Fourier interpolation by densifying the initial \abinitio $\qq$ mesh~\cite{Agapito2018}.
Unfortunately, this assumption does not hold in the presence of such LR behavior: using a denser DFPT mesh localizes the sharp oscillations of the potential in a slightly smaller region around $\qq = \Gamma$ (see red curve in Fig.~\ref{fig:potentials}(c)), but their amplitude is proportional to the discontinuity~\cite{supplemental_prl}. 
In the Supplementary Material, we show that treating the quadrupole interaction is also important when using a Wannier-based interpolation scheme.
This has also been recently confirmed by Jhalani \textit{et al.}~\cite{Jhalani2020}.

\begin{figure}
    \centering
    \includegraphics[clip=0.2cm 0.2cm 0.2cm 0.2cm,width=0.45\textwidth]{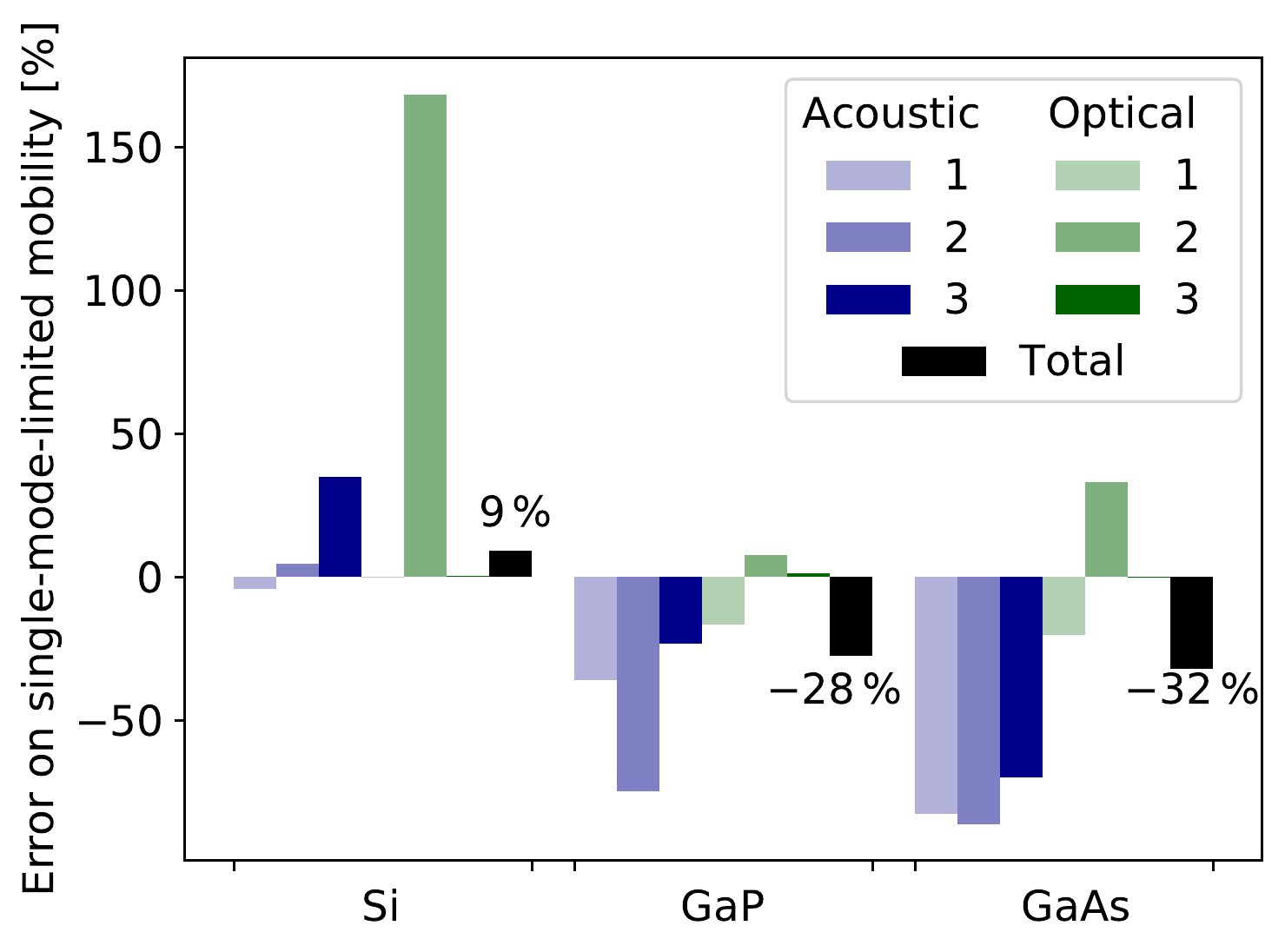}
    \caption{Error on the single-phonon-mode limited mobility when the quadrupole interaction is not correctly treated in Si, GaP and GaAs, for acoustic (blue) and optical (green) modes. The error on the total mobility is given in black.}
    \label{fig:mob_modes}
\end{figure}

At this point, we can quantify the error introduced by these spurious oscillations in the final physical results, and the importance of the quadrupolar interactions.
Because of intravalley transitions, the region around $\Gamma$ (where the LR interactions dominate) is usually one of the most important for the description of e-ph scattering processes~\cite{Brunin2019prb,supplemental_prl}. 
In Fig.~\ref{fig:potentials}, the region around $\Gamma$ relevant for intravalley scattering is represented by gray areas.
An accurate description in this region is crucial for reliable calculations of the phonon-induced electron linewidths,
\begin{equation}
    \begin{split}
        \tau^{-1}_{n\kb} =
                    & 2 \pi \sum_{m,\nu} \int_\BZ \frac{d\qb}{\Omega_\BZ} |\gkkp|^2\\
                    & \times \left[ (n_\qnu + f_{m\kb+\qb})
                                    \delta(\enk - \emkq  + \wnuq) \right.\\
                    & \left. + (n_\qnu + 1 - f_{m\kb+\qb})
                                    \delta(\enk - \emkq  - \wnuq ) \right],
    \end{split}
    \label{eq:tau}
\end{equation}
with $\Omega_\text{BZ}$ the BZ volume, $n_\qnu$ and $f_{m\kb+\qb}$ the Bose-Einstein and Fermi-Dirac
occupation functions and $\enk$ the energy of the electronic state $n\kb$.
These linewidths are needed to compute (phonon-limited) carrier mobilities within the self-energy relaxation time approximation~\cite{Giustino2017,Madsen2018,Ponce2018,Brunin2019prb}:
\begin{equation}
\mu_{e,\alpha\beta} = \frac{-1}{\Omega n_e}
\sum_n \int \frac{d\kb}{\Omega_\BZ} \vnka \vnkb \tau_{n\kb} \left.\frac{\partial f}{\partial\varepsilon}\right|_{\enk},
\label{eq:mobility}
\end{equation}
where $n_e$ is the electron concentration and $\vnka$ is the $\alpha$-th component 
of the velocity operator~\cite{Brunin2019prb}.
Splitting the sum over $\nu$ in Eq.~\eqref{eq:tau}, one can obtain the partial mobility
limited by a single phonon mode.
Figure~\ref{fig:mob_modes} reports the error made on these single-phonon-mode limited mobilities in Si, GaP, and GaAs if quadrupole interactions are not properly treated when interpolating the potentials on the fine 
$\qq$ mesh necessary for convergence~\cite{supplemental_prl}. The initial DFPT $\qq$ meshes are \grid{9} for Si and GaP, and \grid{6} for GaAs. We note that these errors are not significantly reduced by densifying the initial mesh if quadrupole interactions are not treated~\cite{supplemental_prl}.
In Si, the quadrupole interaction mainly affects one of the optical modes. The effect on acoustic modes is much smaller, as expected from the quadrupole acoustic sum rule for nonpolar semiconductors and already stressed by Vogl~\cite{Vogl1976}.
In contrast, in GaP and particularly in GaAs, the acoustic modes are the most affected.
This is also expected since these modes are dominated by the piezoelectric coupling related to acoustic modes~\cite{Vogl1976}. 
When considering the scattering by all phonon modes, the error on the total mobility is around 10\% in Si and goes up to 32\% in GaAs. The larger error in GaAs is linked to its smaller effective mass and the presence of a single valley at $\Gamma$ hence most of the scattering for electrons close to the band edge is intravalley with small $\qq$, see Fig.~\ref{fig:potentials}.
This shows the importance of the quadrupole interaction in understanding the physics of all classes of e-ph scattering. These errors on mobilities can be even more significant than many-body effects (within the transport formalism used in this work). Indeed, 
according to Ponc\'e \textit{et al.} $GW$ corrections of the KS band energies of Si increase the electron mobility by 5\%~\cite{Ponce2018}.
Electron-phonon calculations are usually performed starting from coarse $\qq$ meshes in order to limit
both the number of DFPT computations and the cost of the interpolation itself, which quickly increases 
with the number of $\qq$ points in the initial \abinitio mesh. It is thus not surprising
that this behavior has been largely overlooked so far.

We can also gauge the relative importance of the dynamical quadrupole and the electric-field terms in Eq.~\eqref{eq:LRpot}. 
In nonpolar systems, $\mathbf{Q}_{\kappa\alpha}$ is
the only additional quadrupolar term in the LR scattering potential. 
In the polar systems investigated so far, we observe that 
$\mathbf{Q}_{\kappa\alpha}$ gives a larger contribution to Eq.~\eqref{eq:LRpot}
than the $\cal{E}$ term~\cite{supplemental_prl}.
For instance, ignoring the $\cal{E}$ term changes the electron mobility in GaAs (GaP) by 0.1\% (0.01\%).
In GaAs and GaP, one could speculate that the $\cal{E}$ term is negligible for acoustic modes, where the Fr\"ohlich contribution exactly vanishes.
Additionally, 
$v^{\text{Hxc},\cal{E}}({\bf r})$
oscillates on the scale of the interatomic distances, and couples more weakly to the electronic orbitals than a macroscopic potential shift.
In the ionic limit, for example, one can expect $v^{\text{Hxc},\cal{E}}({\bf r})$ to be 
antisymmetric around each atomic site, 
with a vanishing diagonal matrix element for the corresponding symmetric electronic orbitals. 
Generalizations of the predominance of $\mathbf{Q}_{\kappa\alpha}$ to other systems should be done with care. For example, in many crystals without free Wyckoff parameters (e.g., rocksalt or cubic perovskite) the quadrupole tensor vanishes identically; thus, in such materials the $\cal{E}$ term remains the only source of nonanalytic behavior in the first-order potential besides Fr\"ohlich, and could lead to interesting physics.
Note that this term is absent in the contemporaneous analysis by Jhalani \textit{et al.}~\cite{Jhalani2020}.

We also note that $\mathbf{Q}_{\kappa\alpha}$ introduces
additional dipole-quadrupole and quadrupole-quadrupole terms
at the dynamical matrix level,
which have to be considered for accurate phonon spectra, especially for small $\qq$ where artificial vibrational instabilities may appear~\cite{Stengel2013,Royo2020}. These higher-order terms have also been included in our calculations. 
In Si, GaAs and GaP, they only change the mobility by $\sim$1\%,
which is smaller than the 
effect of the quadrupole potential, but still larger than the effect of the electric-field induced local potential.
However, we cannot exclude that these contributions might be larger for piezoelectric materials, in which spurious instabilities may appear in the phonon dispersion~\cite{Royo2020}.
More details about 
the effects associated with the electric-field induced local potential, to the dipole-quadrupole and the quadrupole-quadrupole interactions in the dynamical matrix are discussed in the Supplemental Material~\cite{supplemental_prl}.


In conclusion, we have included the treatment of quadrupolar fields, beyond the Fr\"ohlich interaction, in the first-principles electron-phonon vertex for semiconductors. By their LR nature, a proper treatment of the quadrupolar fields is necessary for the accurate description of e-ph quantities such as the scattering potential and carrier mobilities. Without treating quadrupolar interactions, large errors from 10 to 30\% can be obtained in evaluating the mobilities of polar and nonpolar semiconductors such as Si, GaP, or GaAs. Taking into account very recent results on GaN and PbTiO$_3$~\cite{Jhalani2020}, we expect these additional contributions to be relevant in a wide variety of materials. For instance, the quadrupolar response is always present in piezoelectric materials~\cite{Royo2019}, where the piezoelectric constants are uniquely defined by the dipole and quadrupole moments~\cite{Martin1972}. These LR physical effects lead to important corrections which can be even more significant than quasiparticle corrections. Since LR macroscopic interactions play a key role in semiconductor physics, we believe that the quest for accurate \abinitio descriptions of phonon-limited carrier mobilities and other e-ph related properties should start from a proper treatment of these physical phenomena.


G.~B., G.-M.~R. and G.~H.~acknowledge financial support from F.R.S.-FNRS.
H.~P.~C.~M. acknowledges financial support from F.R.S.-FNRS through the PDR Grants HTBaSE (T.1071.15).
M.~G., M.~J.~V. and X.~G.~acknowledge financial support from F.R.S.-FNRS through the PDR Grants AIXPHO (T.0238.13) and ALPS (T.0103.19). M.~J.~V thanks FNRS and ULiege for a sabbatical grant in ICN2.
M.~S. and M.~R. acknowledge the support of Ministerio de Economia,
Industria y Competitividad (MINECO-Spain) through
Grants  No.~MAT2016-77100-C2-2-P  and  No.~SEV-2015-0496;
of Generalitat de Catalunya (Grant No.~2017 SGR1506); and
of the European Research Council (ERC) under the European Union's
Horizon 2020 research and innovation program (Grant
Agreement No.~724529).
The present research benefited from computational resources made available on the {Tier-1} 
supercomputer of the F\'ed\'eration Wallonie-Bruxelles, infrastructure funded by the Walloon 
Region under grant agreement n\textsuperscript{o}1117545.

\vspace{2mm}
\textit{Note added} -- Recently, we became aware of a related work
by another group that reaches similar conclusions about the
importance of the dynamical quadrupole term to obtain an
accurate physical description of e-ph interactions~\cite{Jhalani2020,Park2020}.
\vspace{2mm}

\bibliographystyle{apsrev4-1}

%

\end{document}


\title{Electron-phonon 
beyond Fr\"ohlich:
dynamical quadrupoles
in polar and covalent solids \\ Supplemental Material }
\date{\today}
\author{Guillaume Brunin$^1$}
\author{Henrique Pereira Coutada Miranda$^1$}
\author{Matteo Giantomassi$^1$}
\author{Miquel Royo$^2$}
\author{Massimiliano Stengel$^{2,3}$}
\author{Matthieu J. Verstraete$^{4,5}$}
\author{Xavier Gonze$^{1,6}$}
\author{Gian-Marco Rignanese$^1$}
\author{Geoffroy Hautier$^1$}

\affiliation{$^1$UCLouvain, Institute of Condensed Matter and Nanosciences (IMCN), Chemin des \'Etoiles~8, B-1348 Louvain-la-Neuve, Belgium}
\affiliation{$^2$Institut de Ci\`encia de Materials de Barcelona (ICMAB-CSIC), Campus UAB, 08193 Bellaterra, Spain}
\affiliation{$^3$ICREA-Instituci\'o Catalana de Recerca i Estudis Avan\c{c}ats, 08010 Barcelona, Spain}
\affiliation{$^4$NanoMat/Q-Mat/CESAM, Universit{\'e} de Li{\`e}ge (B5), B-4000 Li{\`e}ge, Belgium}
\affiliation{$^5$Catalan Institute of Nanoscience and Nanotechnology (ICN2), Campus UAB, 08193 Bellaterra, Spain}
\affiliation{$^6$Skolkovo Institute of Science and Technology, Moscow, Russia}

\pacs{}

\maketitle

\section{Computational details}

Ground-state and DFPT phonon computations
are performed using norm-conserving pseudopotentials of the Troullier-Martins type~\cite{Troullier1991}, 
in the local-density approximation from Perdew and Wang~\cite{Perdew1992} parametrized by Ceperley and 
Alder~\cite{Ceperley1980}, with a plane-wave kinetic energy cutoff of 20 Ha for Si and 30 Ha for GaAs and GaP. 
For phonon computations, the BZ is sampled using \grid{18} $\Gamma$-centered 
$\kb$-point grids to obtain a good representation of the scattering potentials and a correct convergence 
of the Born effective charges, dielectric and quadrupolar tensors. 

\section{Real-space representation}

Additional information related to Fig.~1 is given in the present section. The goal is to visualize properly, in real space, the potential
created when a Si atom is displaced along the [100] direction of the conventional cell, and to focus on its quadrupolar component.

\begin{figure}
    \centering
    \includegraphics[clip,trim=0cm 0cm 1cm 0cm,width=0.7\textwidth]{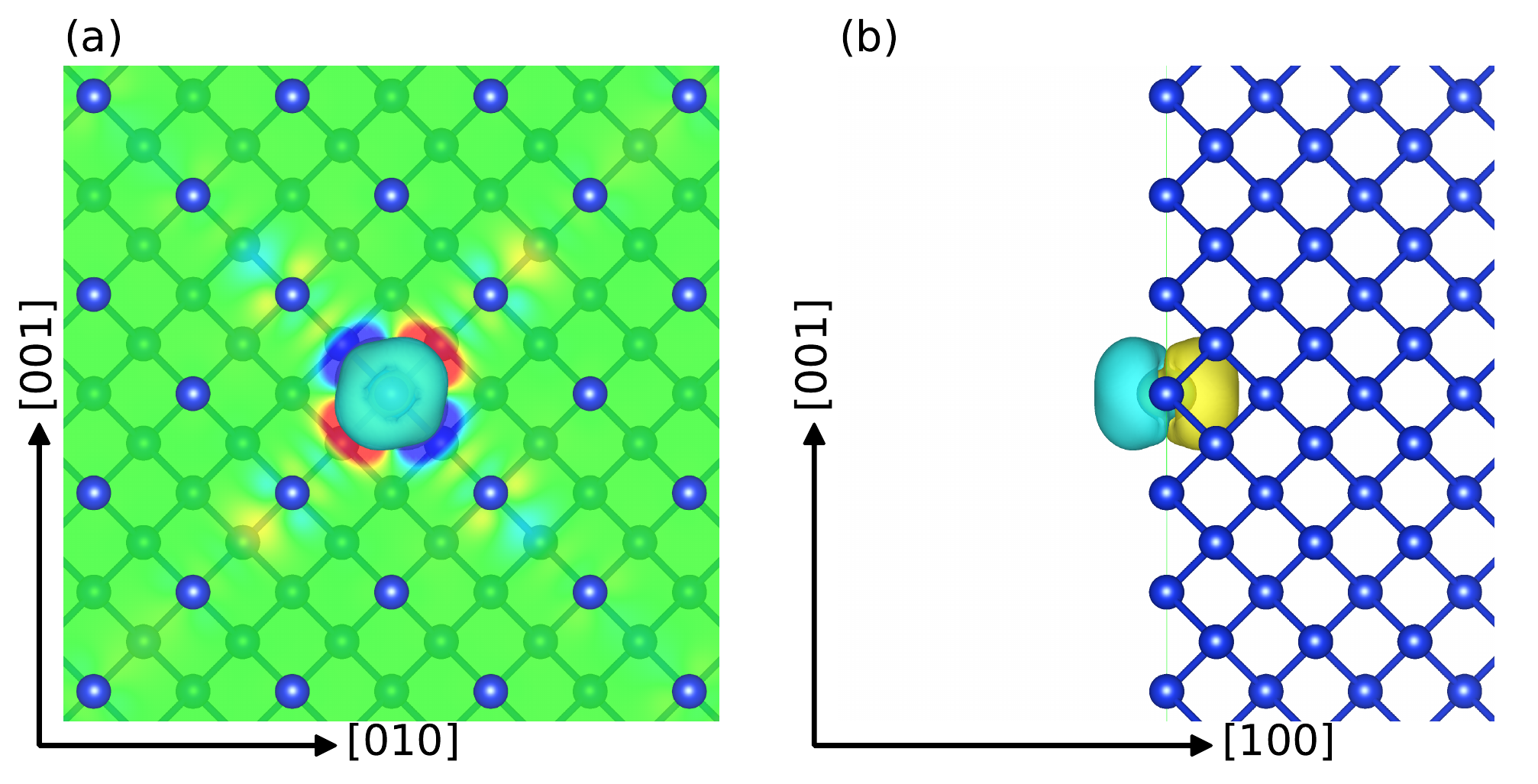}
    \caption{(a) Electronic charge density variation induced by a single
    atomic displacement in Si,
    along the [100] direction and at the center of the conventional cell ({\it perpendicular} to the represented plane).
    Negative (positive) variations are represented in red and yellow (blue and cyan), while values smaller in absolute value than 10\% of the maximum in the plane are represented in green.
    (b) Side view of (a).
    }
\label{fig:DEN}
\end{figure}

The local effect due to the ionic charge displacement and the one due to the electronic charge density combine to give
a zero Born effective charge, hence there is no long-range total dipolar potential. 
The quadrupole survives and dominates the long-range behavior of the potential. 
However, even if the Born effective charge vanishes, the potentials induced {\it separately} by
the ionic displacement and by the electronic
charge density change have a large long-range dipole component, only their sum vanish.
Moreover, in pseudopotential calculations,
the meaning of the potential inside the cut-off radius is obscure.
Hence, special care must be taken to visualize
properly the resulting potential. 
Since the ionic (pseudo)potential is spherically symmetric ($\ell$=0), the corresponding change due to [100] atomic displacement has only an $\ell$=1 (dipole) component. 
Such change vanish in the plane perpendicular to [100]. Similarly, in the plane normal to [100], the $\ell$=1 dipole component associated
to the charge density change vanishes. 
Thus, the quadrupolar change of potential dominates
in this plane, and clearly appears in Fig.~1.
In all the other regions of space, the local $\ell$=1 dipole component does not vanish.
This is illustrated in Figs.~\ref{fig:DEN} (density) and \ref{fig:POT} (potential). 

\begin{figure}
    \centering
    \includegraphics[clip,trim=0cm 0cm 1cm 0cm,width=0.7\textwidth]{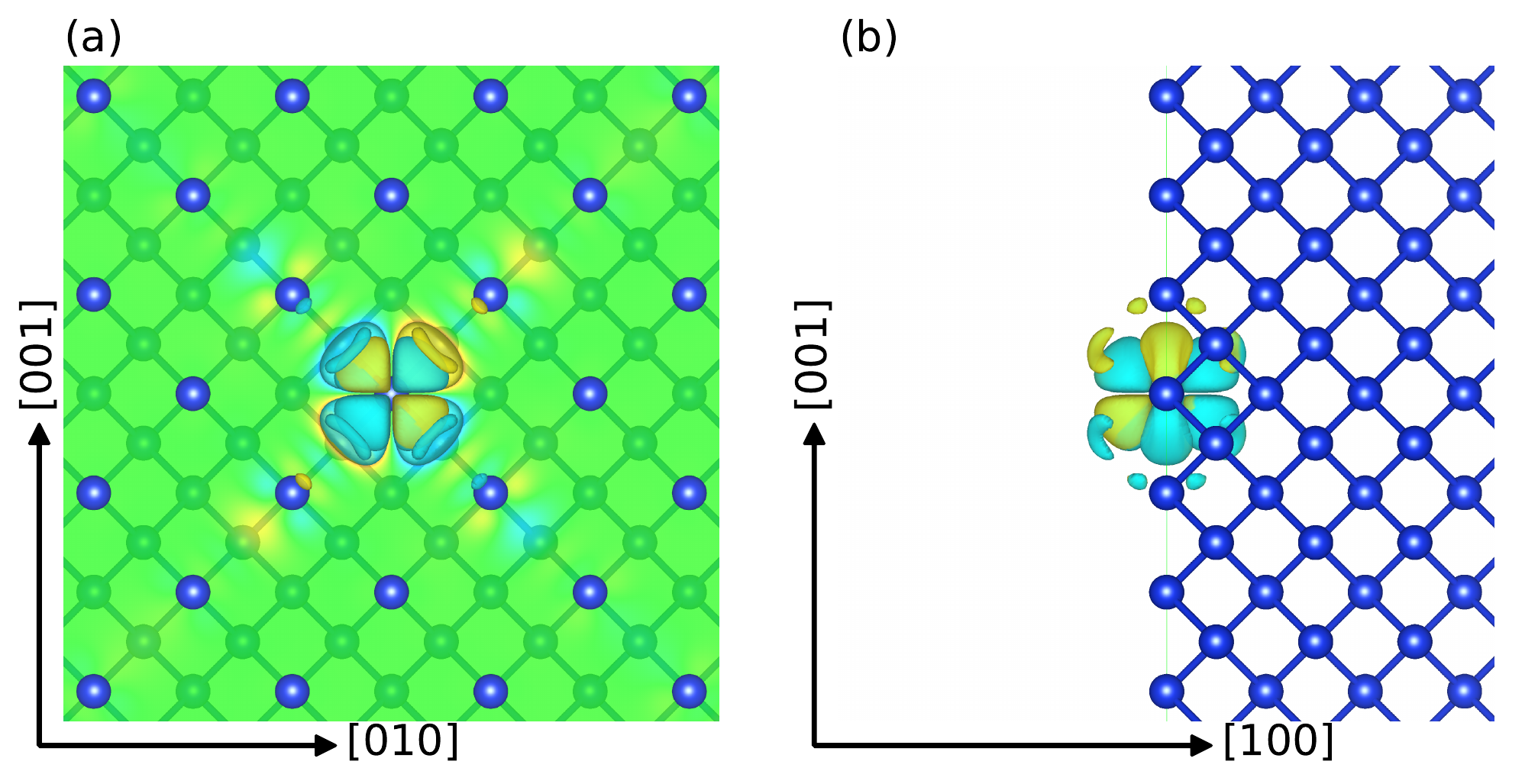}
    \caption{(a) Even part of the electronic charge density variation induced by a single
    atomic displacement in Si,
    along the [100] direction and at the center of the conventional cell ({\it perpendicular} to the represented plane).
    Negative (positive) variations are represented in red and yellow (blue and cyan), while values smaller in absolute value than 10\% of the maximum in the plane are represented in green.
    (b) Side view of (a).
    }
\label{fig:DEN_sym}
\end{figure}

\begin{figure}
    \centering
    \includegraphics[clip,trim=0cm 0cm 1cm 0cm,width=0.7\textwidth]{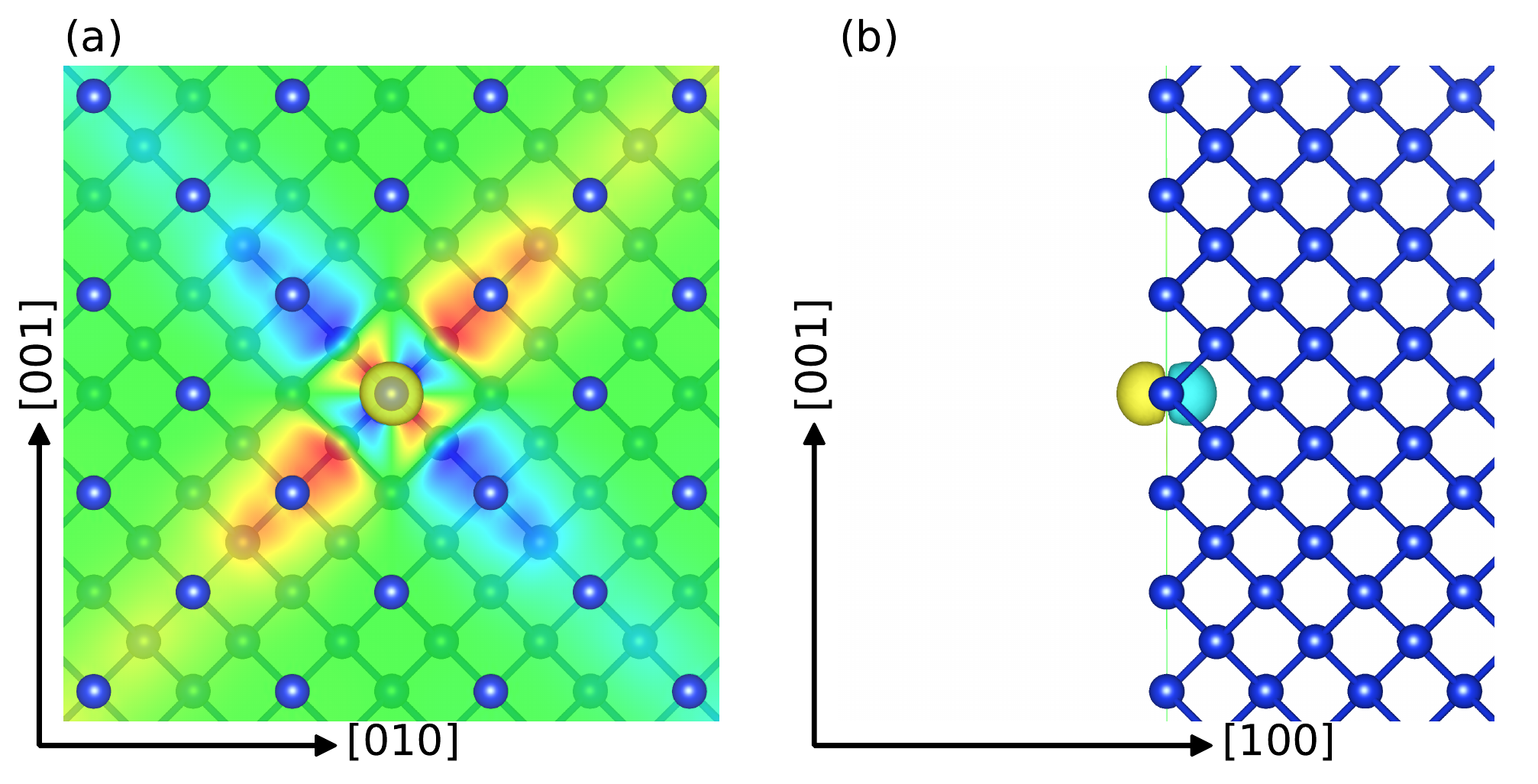}
    \caption{(a) Scattering potential
    induced by a single
    atomic displacement in Si,
    along the [100] direction and at the center of the conventional cell ({\it perpendicular} to the represented plane).
    Negative (positive) variations are represented in red and yellow (blue and cyan), while values smaller in absolute value than 10\% of the maximum in the plane are represented in green.
    (b) Side view of (a).
    }
\label{fig:POT}
\end{figure}

\begin{figure}
    \centering
    \includegraphics[clip,trim=0cm 0cm 1cm 0cm,width=0.7\textwidth]{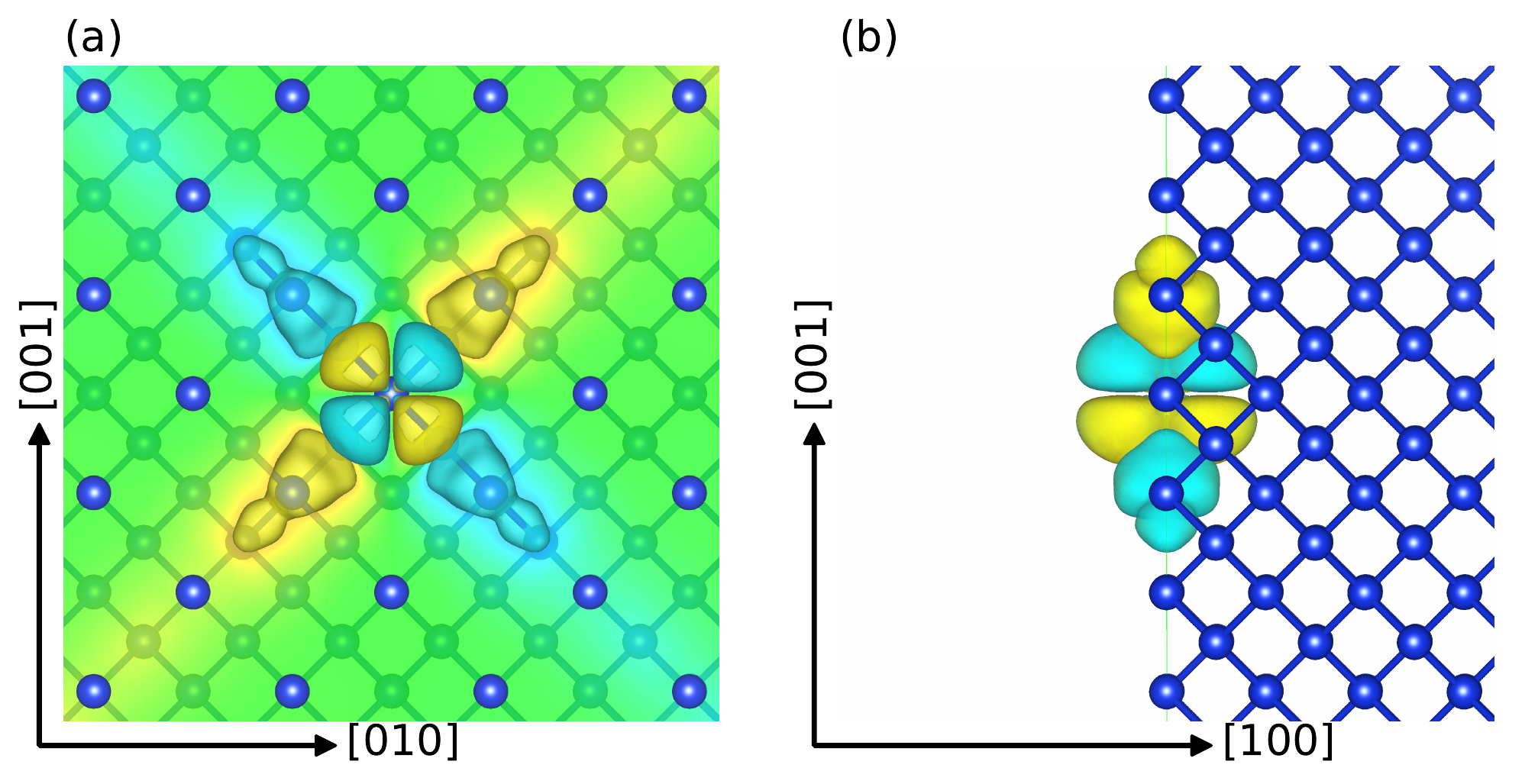}
    \caption{(a) Even part of the scattering potential
    induced by a single
    atomic displacement in Si,
    along the [100] direction and at the center of the conventional cell ({\it perpendicular} to the represented plane).
    Negative (positive) variations are represented in red and yellow (blue and cyan), while values smaller in absolute value than 10\% of the maximum in the plane are represented in green.
    (b) Side view of (a).
    }
\label{fig:POT_sym}
\end{figure}

The $\ell$=1 dipole component can however be eliminated, for the purpose of visualisation,
by {\it projecting out the odd component} of the potential.
For this purpose, the charge density change and scattering potential is decomposed into odd ($o$) and even ($e$) contributions:
\begin{equation}
    \Delta n(\rr) = \Delta n_e(\rr) + \Delta n_o(\rr)
\end{equation}
with
\begin{equation}
    \Delta n_e(\rr) = \frac{\Delta n(\rr) + \Delta n(-\rr)}{2}
\end{equation}
\begin{equation}
    \Delta n_o(\rr) = \frac{\Delta n(\rr) - \Delta n(-\rr)}{2}.
\end{equation}
The same can be written for the real-space scattering potential. The even parts of the charge density change and the scattering potential are represented in Figs.~\ref{fig:DEN_sym} and \ref{fig:POT_sym}. 
The even part of the density change presents characteristics similar to $d$ orbitals, as expected from quadrupolar interactions.

\section{Contribution of the electric-field perturbation to the long-range potentials}

\begin{figure}
    \centering
    \includegraphics[clip,trim=0.2cm 0.2cm 0.2cm 0.2cm,width=.5\textwidth]{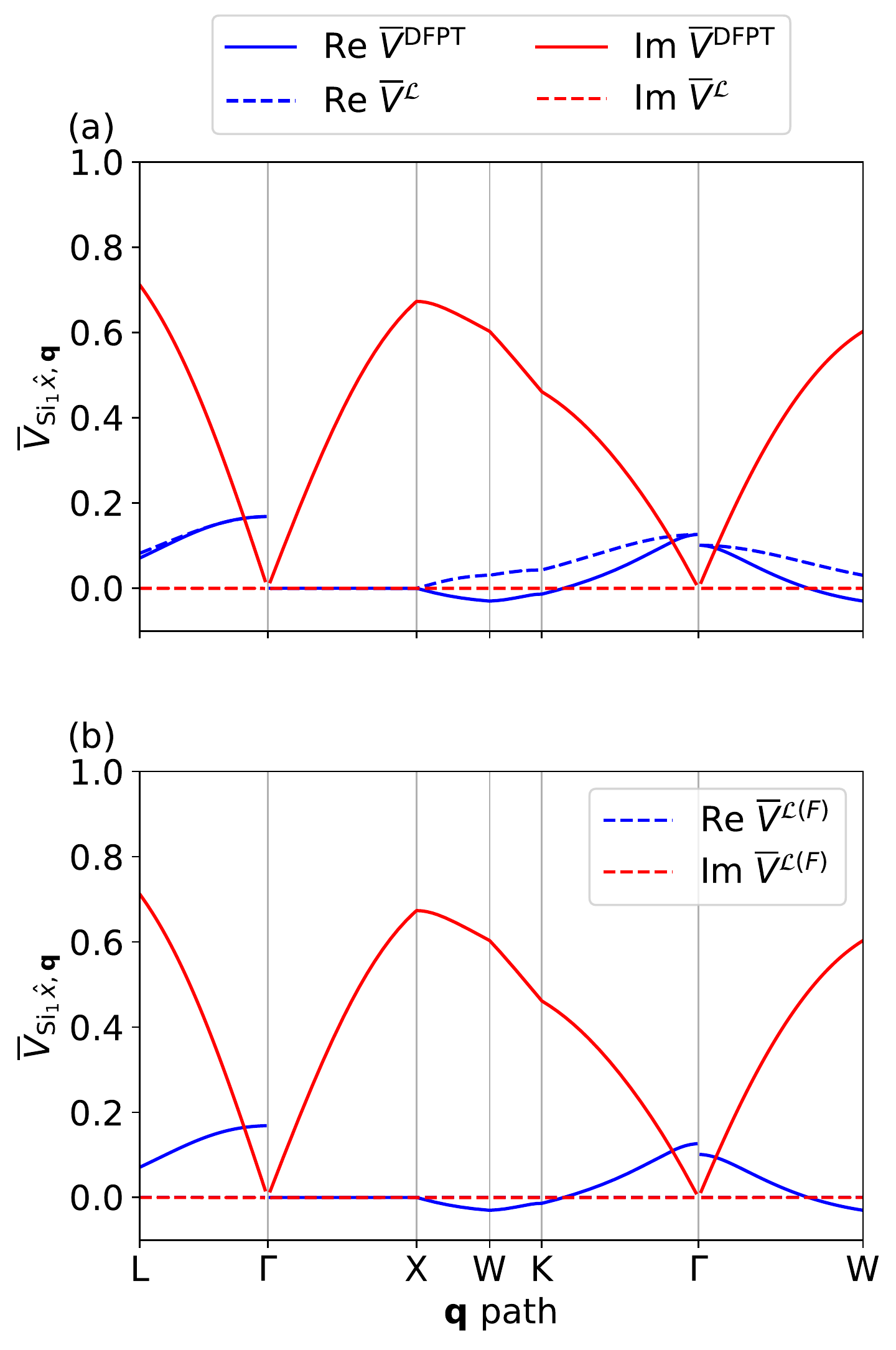}
    \caption{(a) 
    Average over the unit cell of the
    $\hat{x}$ component of the DFPT (full lines) and long-range (dashed) e-ph potentials in Si, 
    for the first atom located at (0,0,0). 
    (b) Same as (a) but with the Fr\"ohlich interaction only (equivalent to zero in Si).}
    \label{fig:Si_Si1_avg_model}
\end{figure}

\begin{figure}
    \centering
    \includegraphics[clip,trim=0.2cm 0.2cm 0.2cm 0.2cm,width=.5\textwidth]{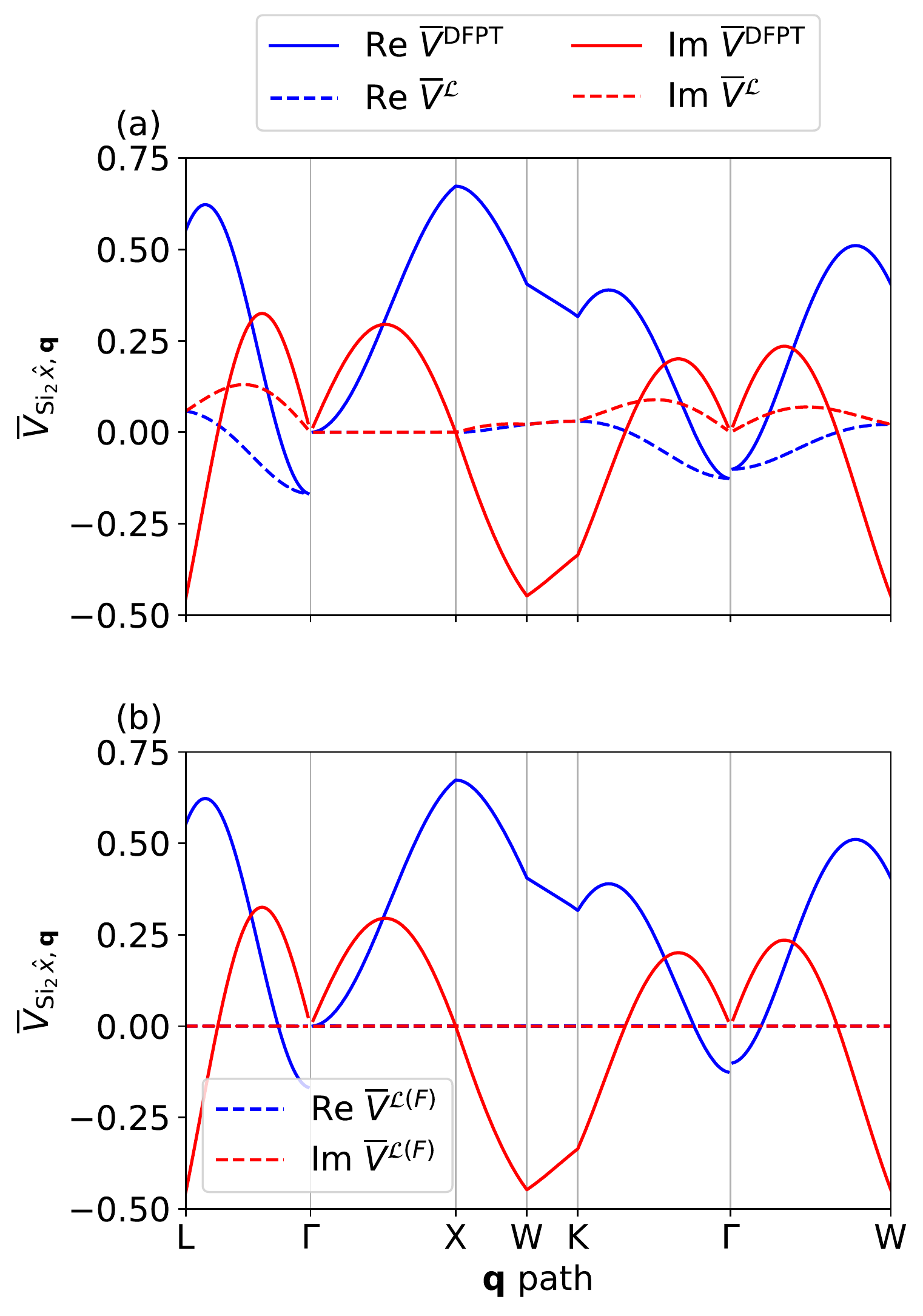}
    \caption{(a) 
    Average over the unit cell of the
    $\hat{x}$ component of the DFPT (full lines) and long-range (dashed) e-ph potentials in Si, 
    for the second atom 
    located at (1/4,1/4,1/4). 
    (b) Same as (a) but with the Fr\"ohlich interaction only (equivalent to zero in Si).}
    \label{fig:Si_Si2_avg_model}
\end{figure}

\begin{figure}
    \centering
    \includegraphics[clip,trim=0.2cm 0.2cm 0.2cm 0.2cm,height=.92\textheight]{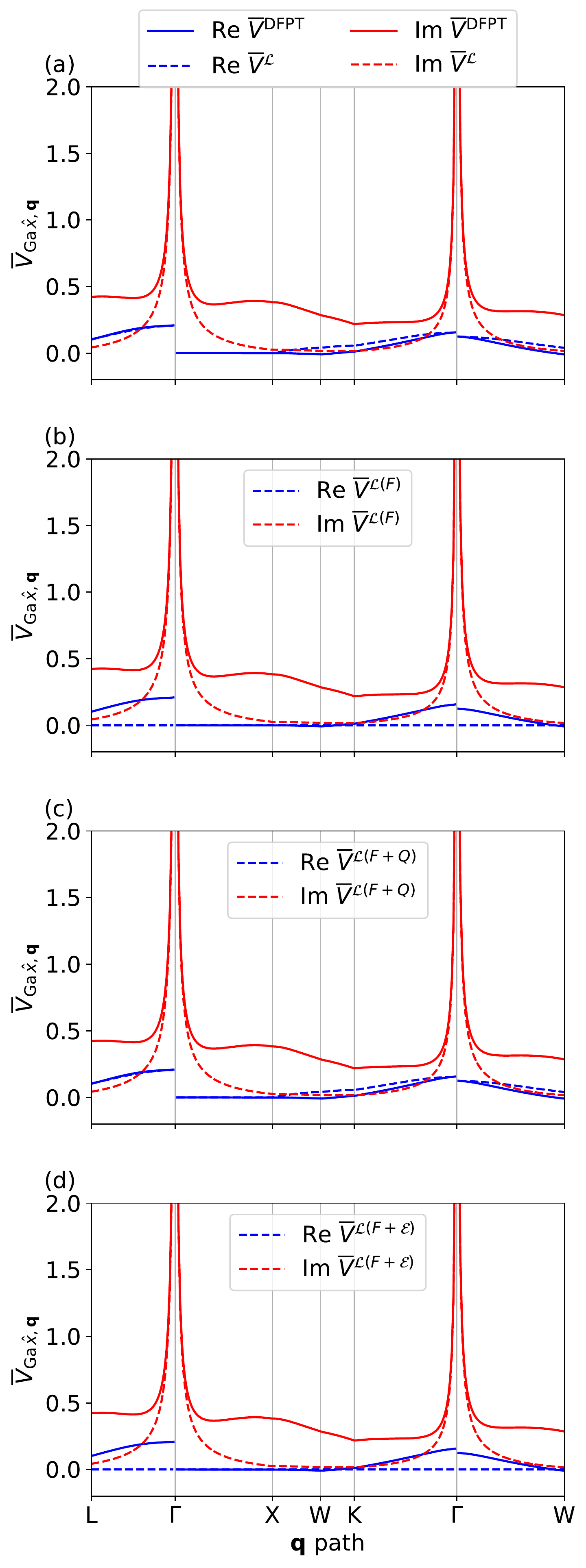}
    \caption{(a) 
    Average over the unit cell of the 
    $\hat{x}$ component of the DFPT (full lines) and long-range (dashed) e-ph potentials in GaAs 
    for the Ga atom 
    located at (0,0,0).
    (b), (c) and (d) show the decomposition of the different contributions to the long-range model.}
    \label{fig:GaAs_Ga_avg_model}
\end{figure}

\begin{figure}
    \centering
    \includegraphics[clip,trim=0.2cm 0.2cm 0.2cm 0.2cm,height=.92\textheight]{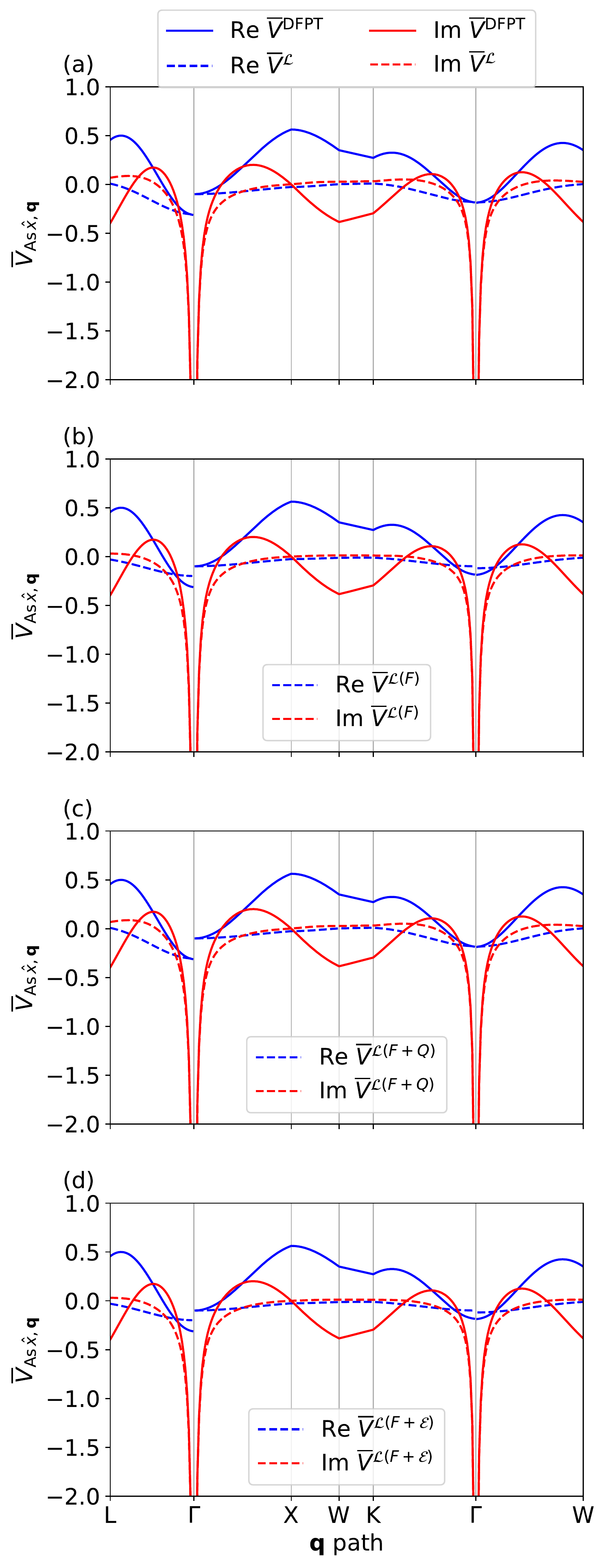}
    \caption{(a) Average over the unit cell of the $\hat{x}$ component of the DFPT (full lines) and long-range (dashed) e-ph potentials in GaAs, 
    for the As atom 
    located at (1/4,1/4,1/4).
    (b), (c) and (d) show the decomposition of the different contributions to the long-range model.}
    \label{fig:GaAs_As_avg_model}
\end{figure}

In this section, we discuss the relative importance of the different terms in the LR model with particular emphasis on the effect associated to the response to the electric field $\cal{E}$.
Figures~\ref{fig:Si_Si1_avg_model}--\ref{fig:GaAs_As_avg_model} show the average over the unit cell of the long-range model including different contributions: $V^{\mathcal{L}}$ includes Fr\"ohlich, dynamical quadrupoles  $\mathbf{Q}_{\kappa\alpha}$ and electric-field contributions, $V^{\mathcal{L}(F)}$ includes only the Fr\"ohlich-like part, $V^{\mathcal{L}(F+Q)}$ includes all but the electric-field perturbation while $V^{\mathcal{L}(F+\mathcal{E})}$ includes all but the dynamical quadrupoles. The potentials are plotted along a $\qq$ path, for a given atom (Ga, As, or Si) displaced along the first reduced direction ($\hat{x}$). 
The average of the model is compared to the average of the \emph{exact} DFPT potentials obtained along the same $\qq$ path. 
Comparing the different subplots produced with different models allows one to have a qualitative understanding of the importance of the different terms and of their effect on the Fourier interpolation of the short-range part.
More specifically, the plots show that the Fr\"ohlich dipolar term alone is not able to capture the jump discontinuities around the $\Gamma$ point, and that quadrupolar fields are necessary (in particular the term stemming from $\mathbf{Q}_{\kappa\alpha}$ as the contribution given by the electric-field perturbation 
is much smaller).

\begin{figure}
    \centering
    \includegraphics[clip,trim=0.2cm 0.2cm 0.2cm 0.2cm,width=.6\textwidth]{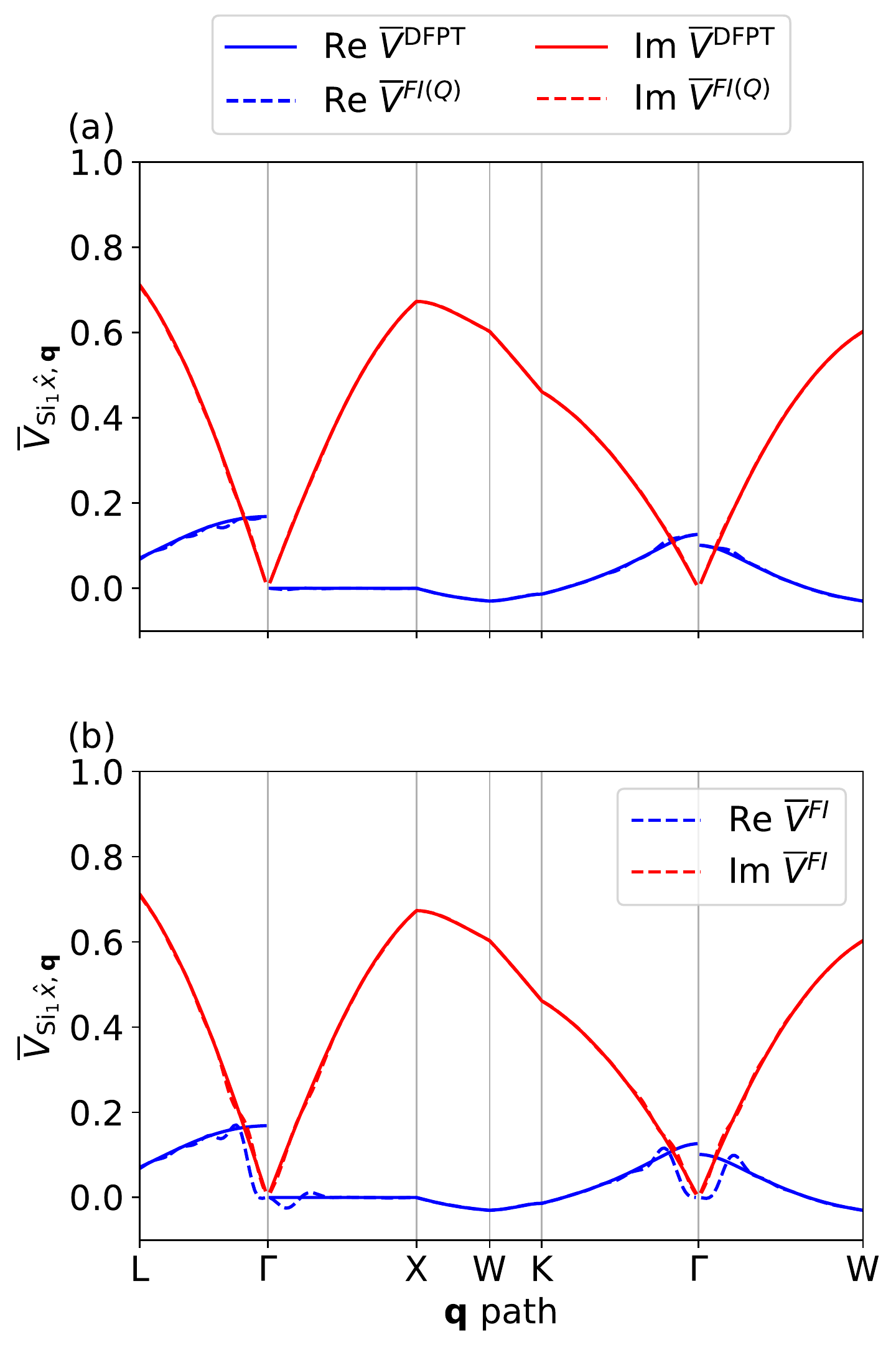}
    \caption{(a) Average over of the unit cell of the $\hat{x}$ component of the DFPT (full lines) and Fourier-interpolated (dashed) e-ph potentials in Si,
    for the first Si atom 
    located at (0,0,0). 
    (b) Same as (a) but with the Fr\"ohlich interaction only (equivalent to zero in Si).}
    \label{fig:Si_Si1_avg_interp}
\end{figure}

\begin{figure}
    \centering
    \includegraphics[clip,trim=0.2cm 0.2cm 0.2cm 0.2cm,width=.6\textwidth]{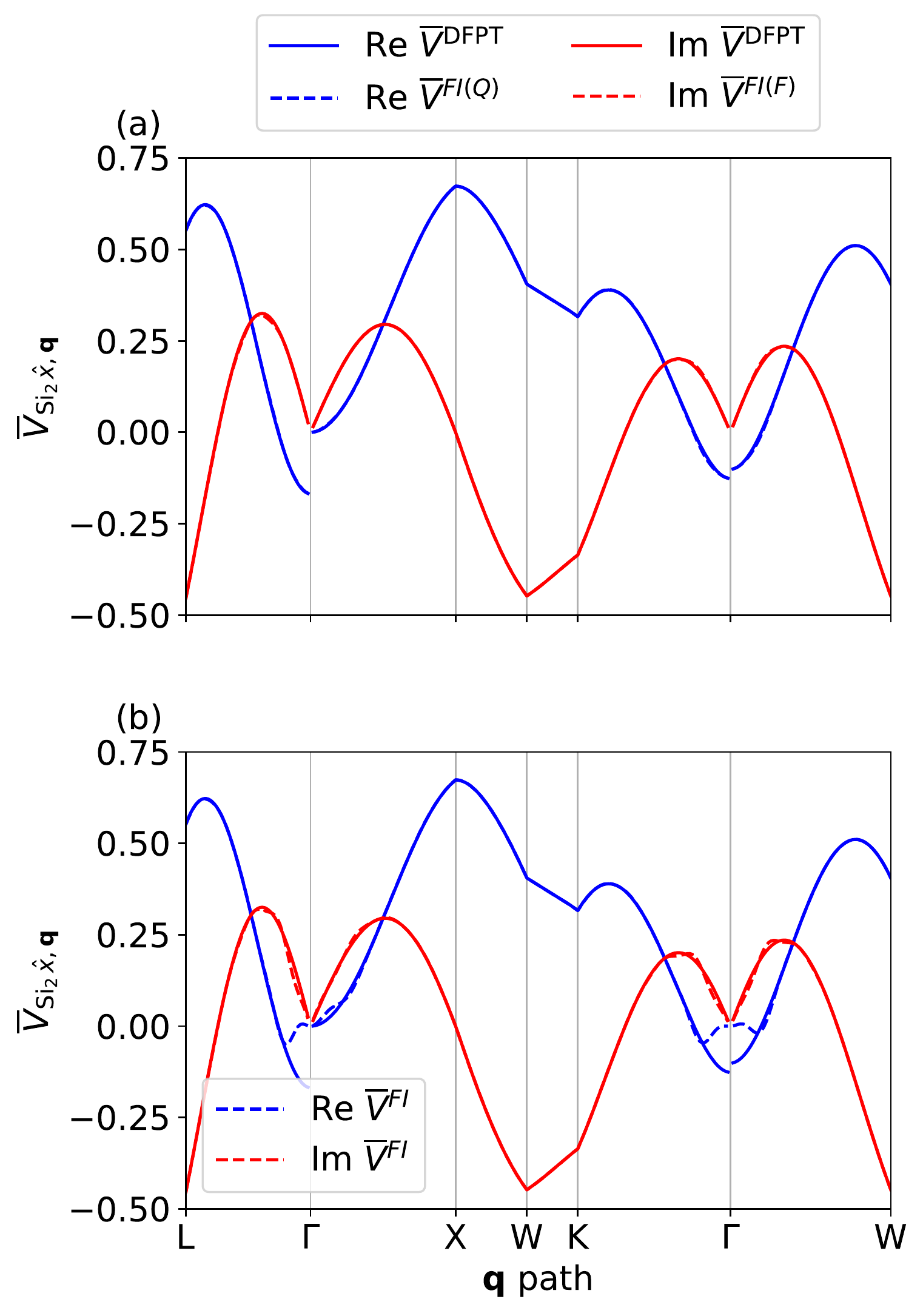}
    \caption{(a) Average over the unit cell of the $\hat{x}$ component of the DFPT (full lines) and Fourier-interpolated (dashed) e-ph potentials in Si, 
    for the second atom 
    located at (1/4,1/4,1/4). 
    (b) Same as (a) but with the Fr\"ohlich interaction only (equivalent to zero in Si).}
    \label{fig:Si_Si2_avg_interp}
\end{figure}

\begin{figure}
    \centering
    \includegraphics[clip,trim=0.2cm 0.2cm 0.2cm 0.2cm,height=.92\textheight]{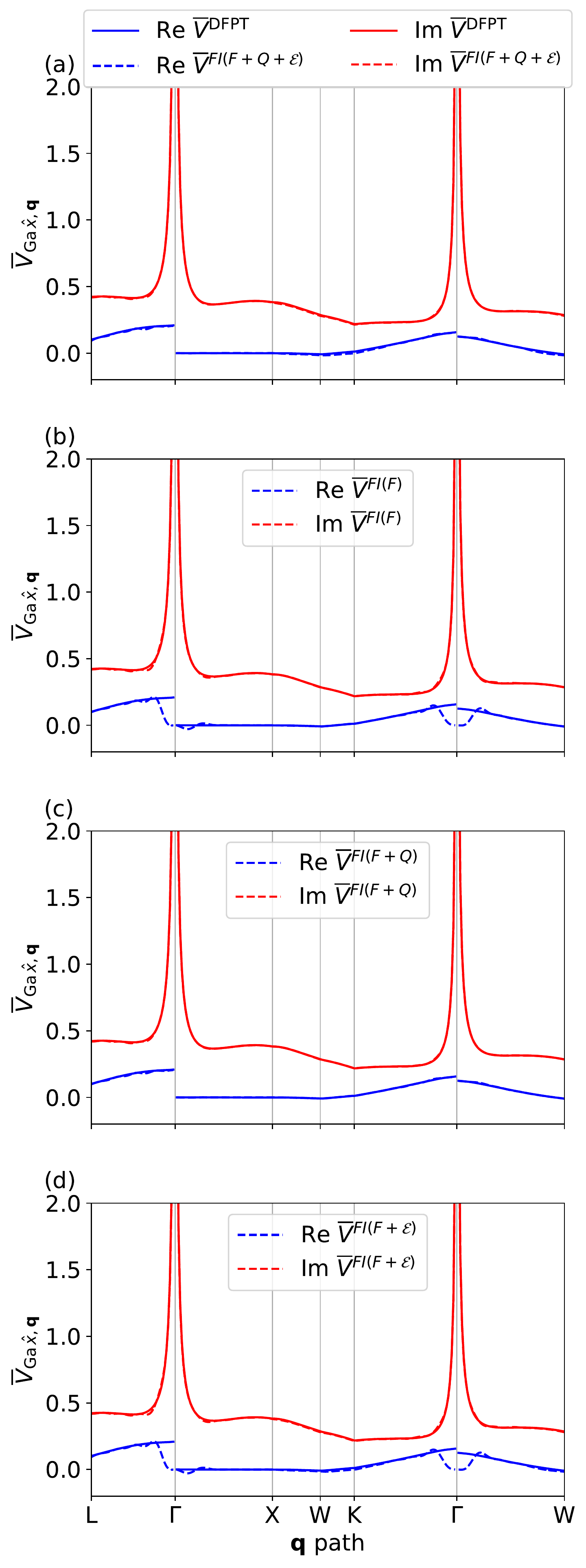}
    \caption{(a) Average over the unit cell of the $\hat{x}$ component of the DFPT (full lines) and Fourier-interpolated (dashed) e-ph potentials in GaAs
    for the Ga atom 
    located at (0,0,0). 
    (b), (c) and (d) show the decomposition of the different contributions to the long-range model.}
    \label{fig:GaAs_Ga_avg_interp}
\end{figure}

\begin{figure}
    \centering
    \includegraphics[clip,trim=0.2cm 0.2cm 0.2cm 0.2cm,height=.92\textheight]{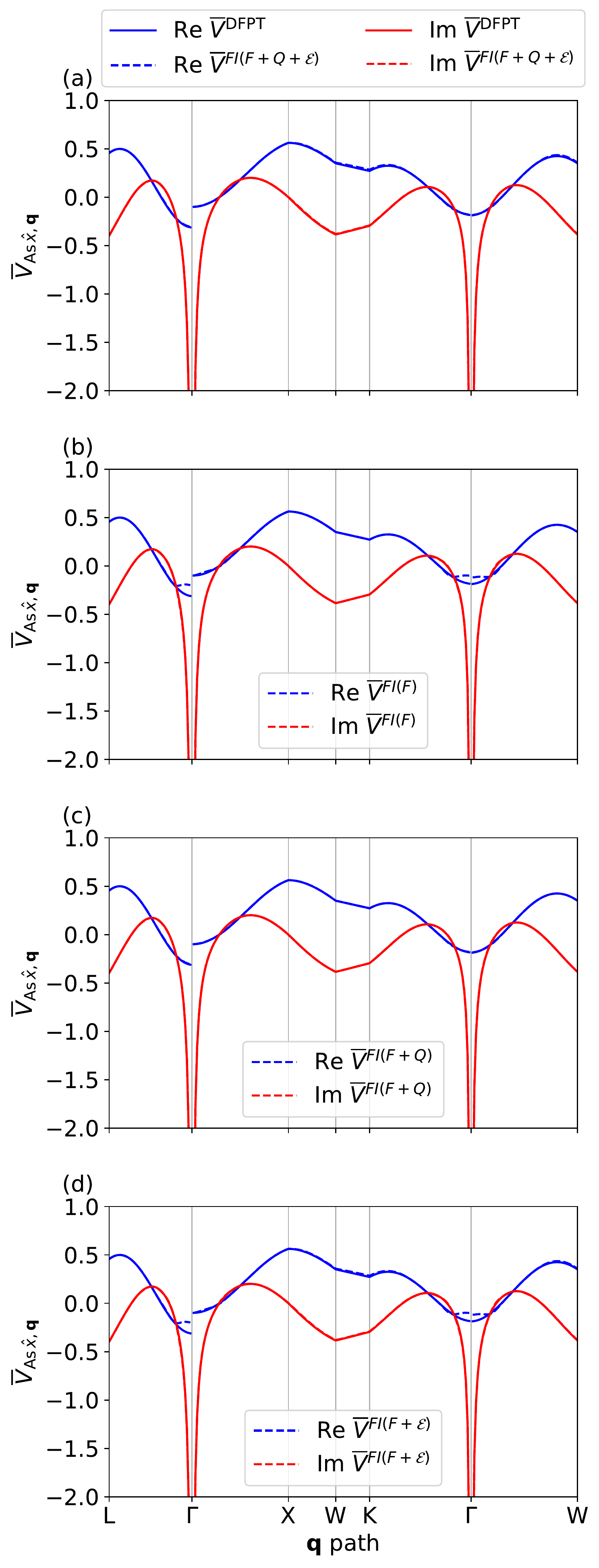}
    \caption{(a) Average over the unit cell of the $\hat{x}$ component of the DFPT (full lines) and Fourier-interpolated (dashed) e-ph potentials in GaAs, 
    for the As atom 
    located at (1/4,1/4,1/4). 
    (b), (c) and (d) show the decomposition of the different contributions to the long-range model.}
    \label{fig:GaAs_As_avg_interp}
\end{figure}

Figures~\ref{fig:Si_Si1_avg_interp}--\ref{fig:GaAs_As_avg_interp} show the average over the unit cell of the periodic part of the Fourier interpolated potentials including the same contributions as in Figs.~\ref{fig:Si_Si1_avg_model}--\ref{fig:GaAs_As_avg_model}, and compared again to the \emph{exact} DFPT potentials. 
These figures reveal that, without $\mathbf{Q}_{\kappa\alpha}$, 
the interpolated scattering potentials are affected by 
unphysical Gibbs oscillations around $\Gamma$
because the discontinuity in $\qq$ space is not removed when a model containing only the dipolar term is subtracted from the initial DFPT potentials before computing the Fourier transform. 

\begin{figure}
    \centering
    \includegraphics[clip,trim=0.2cm 0.2cm 0.2cm 0.2cm,height=.92\textheight]{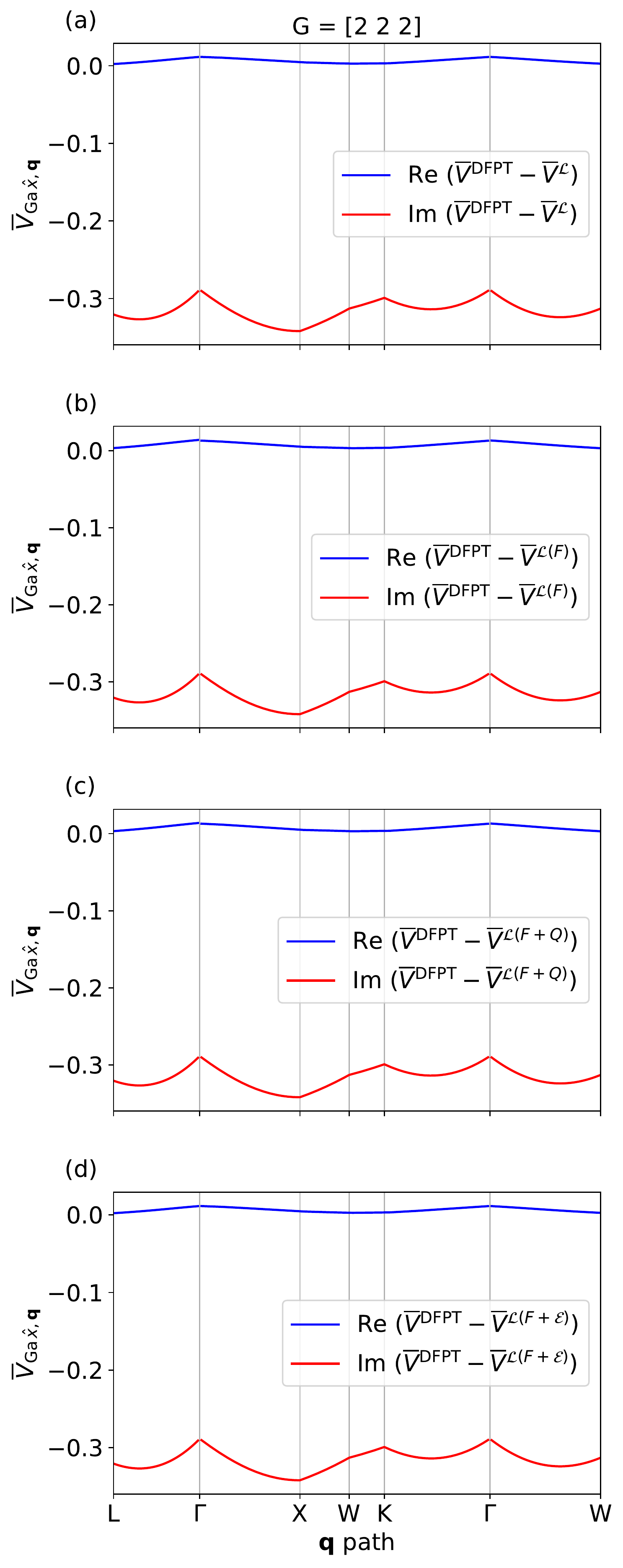}
    \caption{(a) Fourier component of the $\hat{x}$ component of the DFPT (full lines) and long-range (dashed) e-ph potentials in GaAs, 
    for the Ga atom 
    located at (0,0,0). 
    (b), (c) and (d) show the decomposition of the different contributions to the long-range model.
    Only a specific $\Gb$ component of the potential is shown.}
    \label{fig:GaAs_Ga_gsmall}
\end{figure}

\begin{figure}
    \centering
    \includegraphics[clip,trim=0.2cm 0.2cm 0.2cm 0.2cm,height=.92\textheight]{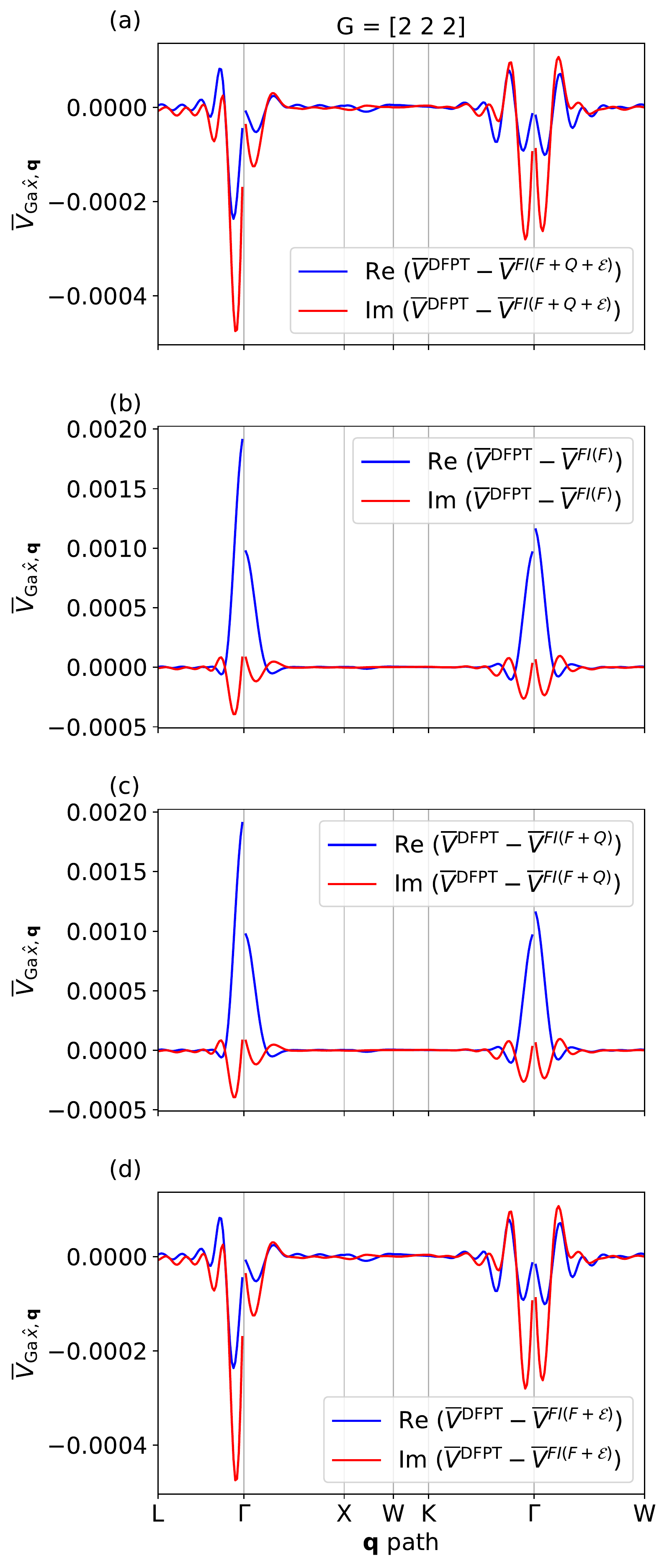}
    \caption{(a) Fourier component of the $\hat{x}$ component of the DFPT (full lines) and Fourier-interpolated (dashed) e-ph potentials in GaAs, 
    for the Ga atom 
    located at (0,0,0). 
    (b), (c) and (d) show the decomposition of the different contributions to the long-range model.
    Only a specific $\Gb$ component of the potential is shown.}
    \label{fig:GaAs_Ga_gsmall_interp}
\end{figure}

As mentioned before, 
the term associated to $\cal{E}$ plays a very minor role if we focus on the average of the potential over the unit cell that corresponds to the $\GG = 0$ Fourier component.
To appreciate the effect of the electric-field term, we need to focus on the $\qq$ dependence of the 
Fourier components of the scattering potentials
\begin{equation}
\label{eq:vqg}
\bar{V}_{\kappa\alpha,\qb}(\Gb) =
\dfrac{1}{\Omega}\int_\Omega \dd\rr\,V_{\kappa\alpha,\qb}(\rb) e^{-i(\qG)\cdot\rr},
\end{equation}
for small $\GG \neq 0$.
In our tests we found that, for particular $\GG$ vectors, the quantity in Eq.~\ref{eq:vqg} as a function of $\qq$ presents (small) jump discontinuities for $\qq \rightarrow 0$ and that the discontinuity is better described when the  $\cal{E}$ term is included in the LR model.
The results are summarized in Figs.~\ref{fig:GaAs_Ga_gsmall} and \ref{fig:GaAs_Ga_gsmall_interp}.
%
We performed the same analysis for GaP (not shown) and found very similar behavior. 
At the level of the mobility, 
the inclusion of the electric-field perturbation in the LR model changes the final results by 0.1\% in GaAs and 0.01\% in GaP when compared to calculations in which only dipoles and $\mathbf{Q}_{\kappa\alpha}$ are included.
These results corroborate our affirmation done in the main text 
about the predominance of the dynamical quadrupoles over the electric field term in the case of GaAs and GaP.
%

\section{Effects of quadrupoles on the matrix elements in Si}

\begin{figure}
    \centering
    \includegraphics[clip,trim=0.2cm 0.2cm 0.2cm 0.2cm,width=.6\textwidth]{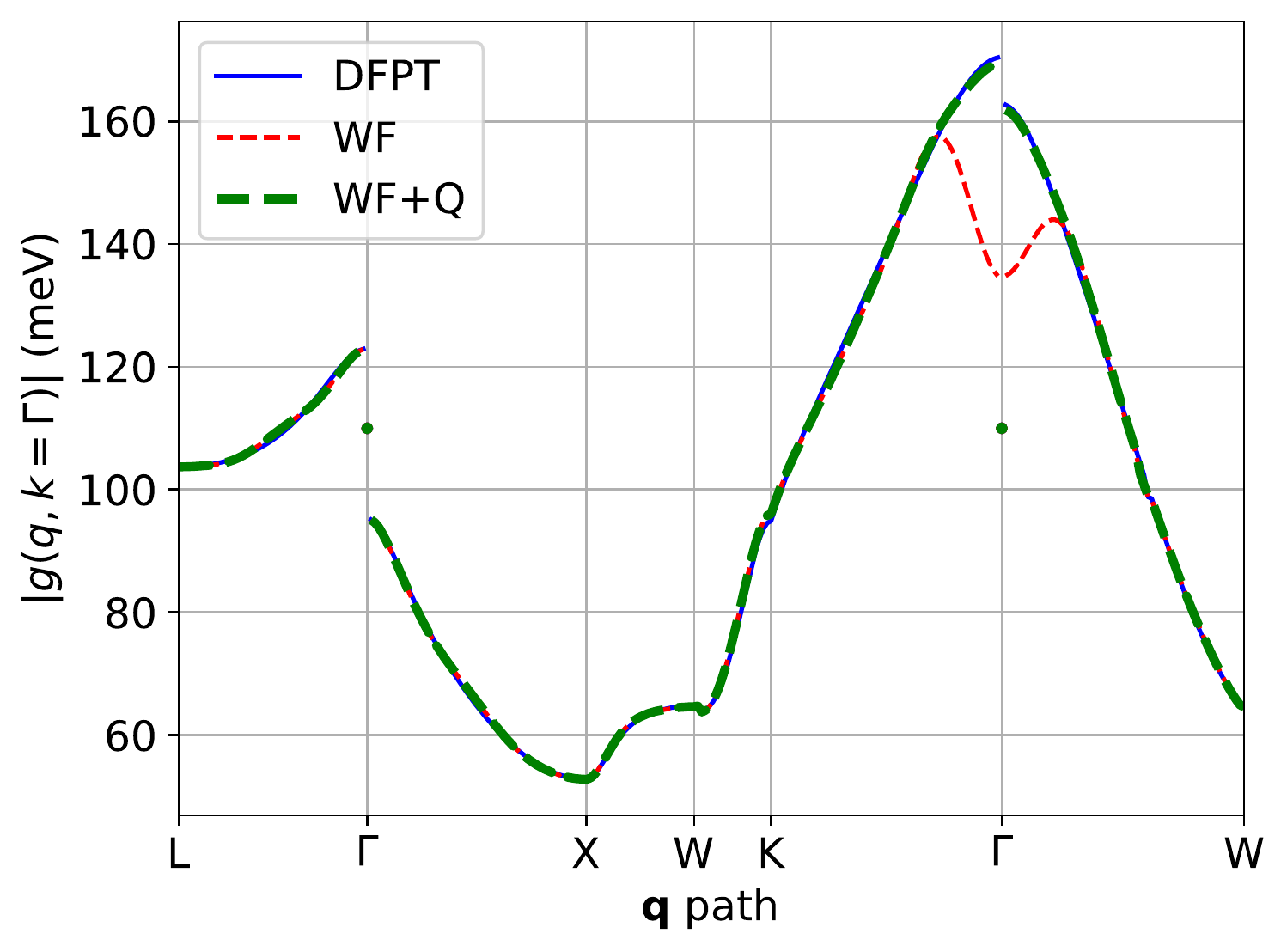}
    \caption{E-ph matrix elements in Si computed with \qe,
    and interpolated using Wannier functions with the standard \epw approach (WF) or
    including the quadrupolar corrections (WF+Q).
    }
    \label{fig:wannier}
\end{figure}

To confirm the important effects of the quadrupoles in Wannier-based interpolation schemes, we modified the EPW code~\cite{Ponce2016}
to include the $\mathbf{Q}_{\kappa\alpha}$ terms in the interpolation of the e-ph matrix elements 
in the Wannier representation in Si. These results have been obtained with Quantum Espresso (\qe) and \epw. We used the Perdew-Burke-Ernzerhof generalized-gradient approximation (GGA) 
for the exchange correlation functional and 
norm-conserving pseudopotentials from the \pseudodojo project~\cite{Vansetten2018, Hamann2013}. We used an energy cutoff of 25 Ry for the plane-wave basis set and a lattice parameter of 5.47~\AA. 
The matrix elements are interpolated starting from coarse $\Gamma$-centered \grid{12} $\kb$-point and \grid{6} $\qb$-point meshes. For the implementation of the quadrupolar corrections 
in \epw, we employed the quadrupole tensors computed by \abinit
as the same computation is not available in \qe. Because of the different pseudopotentials, lattice parameters, and computational settings, we had to adjust the value of $\mathbf{Q}_{\kappa\alpha}$ to better describe the discontinuity, from 13.67 with LDA FHI~\cite{Brunin2019prb} to 15.75 with GGA-PBE.
Figure~\ref{fig:wannier} shows the e-ph
matrix elements connecting the electron state at the valence band maximum at $\kb = \Gamma$
to the other highest states of the valence band through the highest phonon mode at L as in 
Ref.~\cite{Agapito2018}. The DFPT results (blue solid line) and the Wannier-interpolated
results (red and green dashed lines) are compared in this figure. 
The interpolation of the matrix elements using the standard approach 
(\ie without $\mathbf{Q}_{\kappa\alpha}$) 
leads to systematic errors in the region around the $\Gamma$ point
associated to electronic transitions with small momentum transfer.
Including the dynamical quadrupoles in the model 
significantly improves the quality of the interpolant around the $\Gamma$ point.
Similar behavior is also expected in e-ph calculations based on supercells and finite differences~\cite{Chaput2019}.

\section{Effect of dynamical quadrupoles on the interpolated phonon dispersion}

E-ph calculations are quite sensitive to the fine details of the electron and phonon dispersions.
As discussed in more details in Ref.~[15] 
of the main text, an expansion of the dynamical matrix for $\qq \rightarrow 0$ contains additional LR terms beyond the dipole-dipole interaction.
These corrections are supposed to improve the quality and the stability of the Fourier interpolation of the phonon frequencies.
It is therefore interesting to analyze the effect of the $\mathbf{Q}_{\kappa\alpha}$ on the convergence rate of the vibrational spectra for the three systems considered in this work.

\begin{figure}
    \centering
    \includegraphics[clip,trim=0.2cm 0.2cm 0.2cm 0.2cm,width=.9\textwidth]{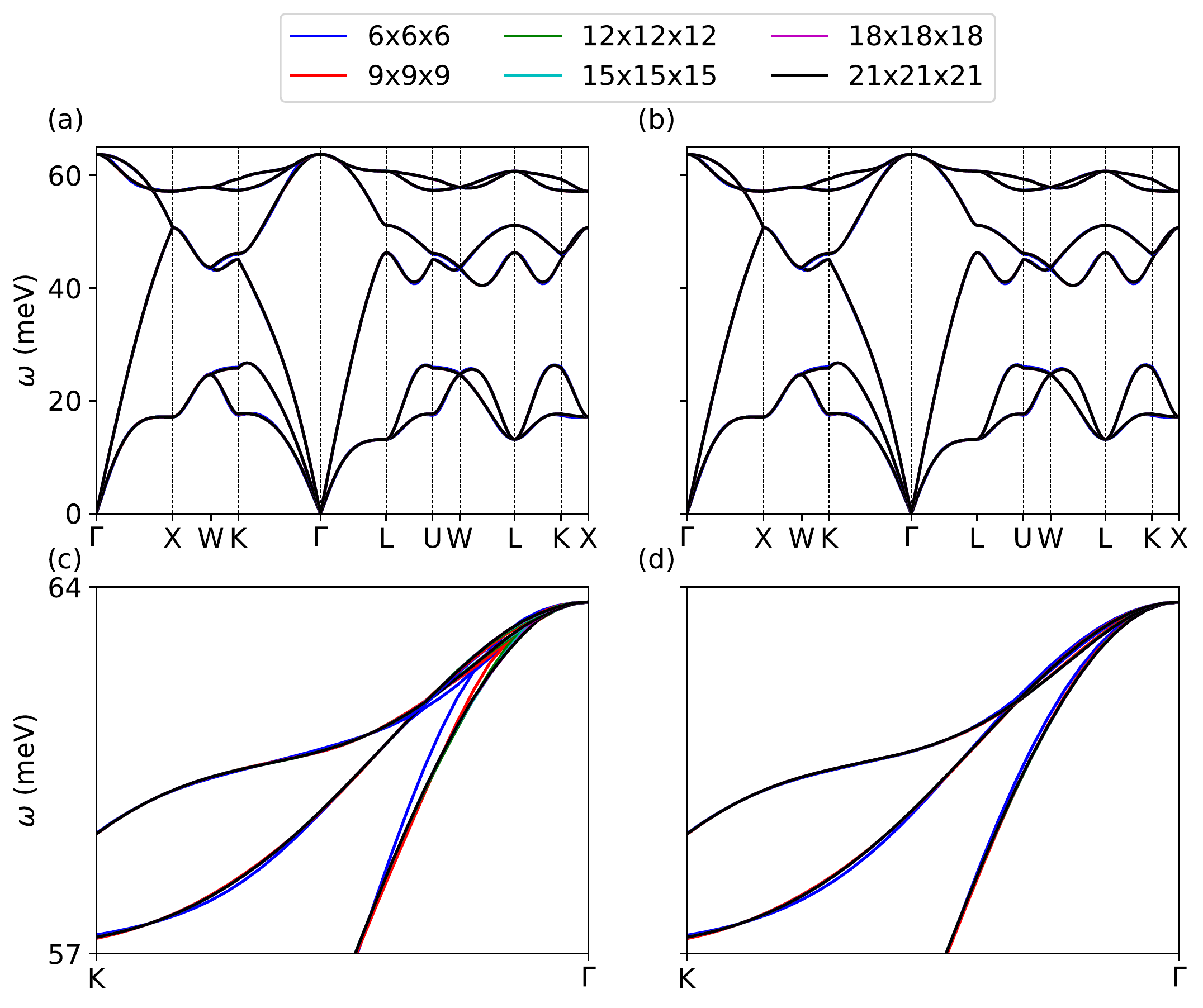}
    \caption{Phonon dispersion in Si without (a) and with (b) the inclusion of dipole-quadrupole and quadrupole-quadrupole interactions in the Fourier interpolation of the dynamical matrix. (c), (d) Zoom over the regions in (a), (b) containing the most important changes.}
    \label{fig:Si_ph_conv}
\end{figure}

\begin{figure}
    \centering
    \includegraphics[clip,trim=0.2cm 0.2cm 0.2cm 0.2cm,width=.9\textwidth]{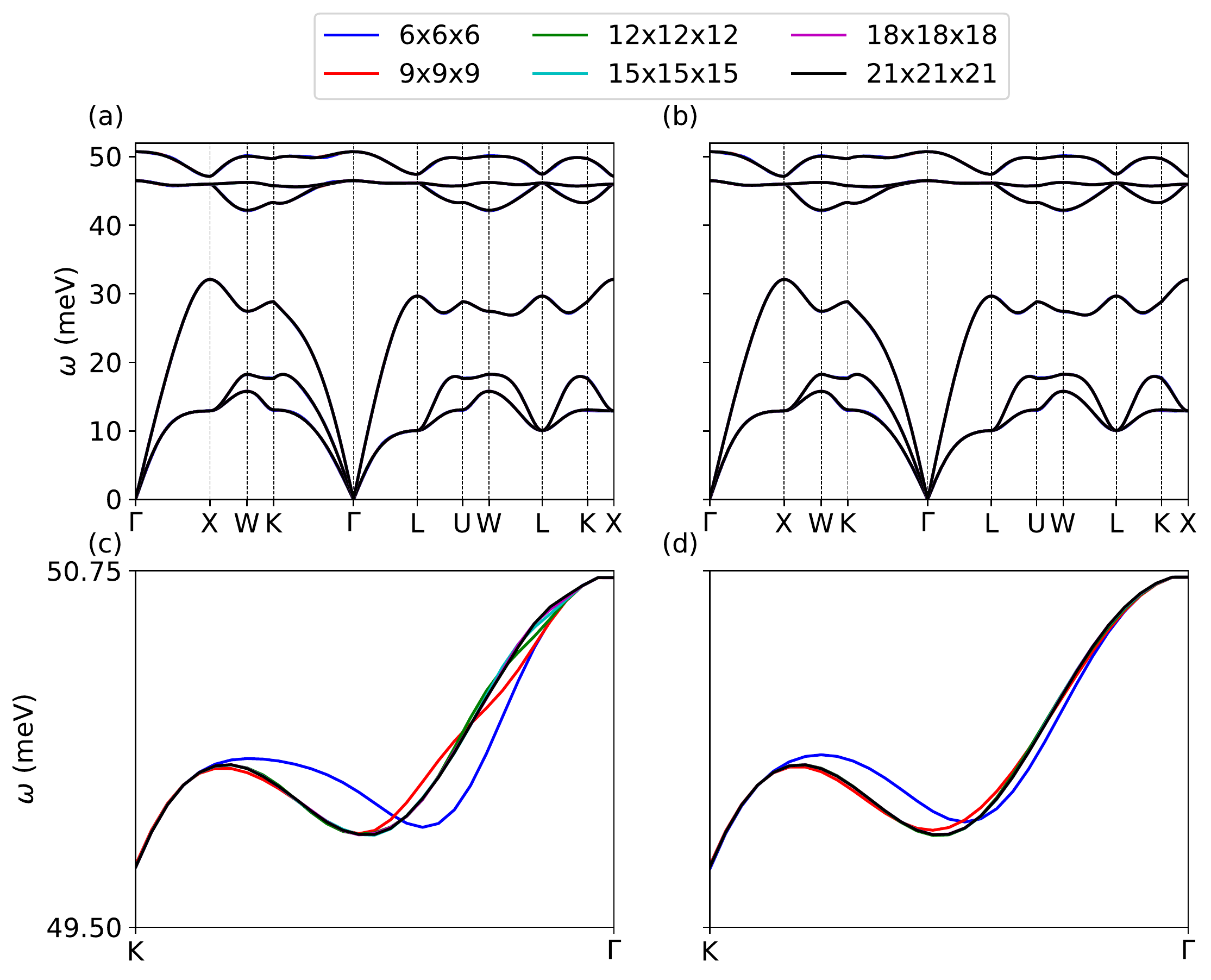}
    \caption{Phonon dispersion in GaP without (a) and with (b) the inclusion of dipole-quadrupole and quadrupole-quadrupole interactions in the Fourier interpolation of the dynamical matrix. (c), (d) Zooms over the regions in (a), (b) containing the most important changes.}
    \label{fig:GaP_ph_conv}
\end{figure}

\begin{figure}
    \centering
    \includegraphics[clip,trim=0.2cm 0.2cm 0.2cm 0.2cm,width=.9\textwidth]{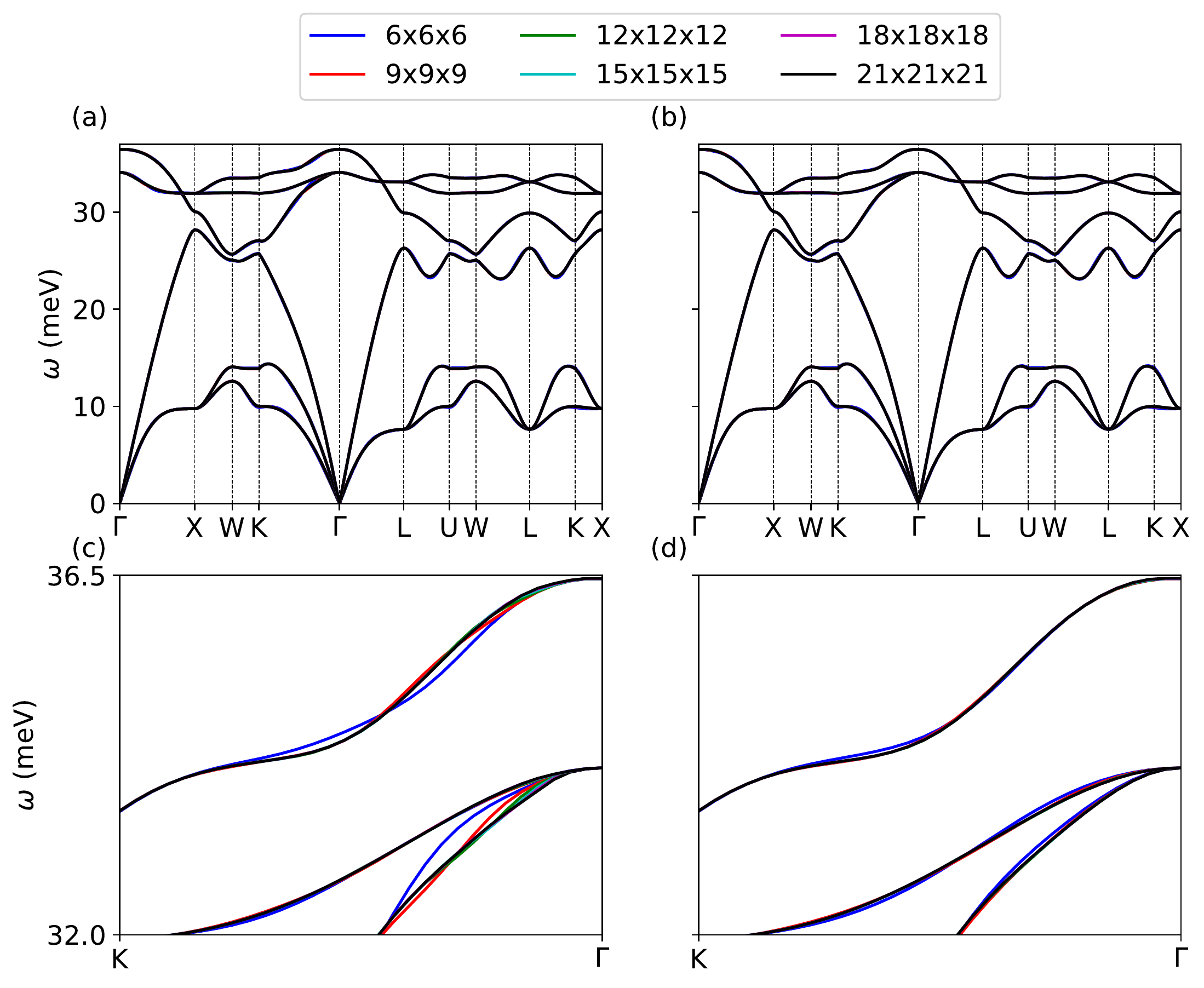}
    \caption{Phonon dispersion in GaAs without (a) and with (b) the inclusion of dipole-quadrupole and quadrupole-quadrupole interactions in the Fourier interpolation of the dynamical matrix. (c), (d) Zoom over the regions in (a), (b) containing the most important changes.}
    \label{fig:GaAs_ph_conv}
\end{figure}

Figure~\ref{fig:Si_ph_conv} shows the influence of the dipole-quadrupole and quadrupole-quadrupole interactions on the Fourier interpolation of the dynamical matrix in Si.
The effects are small on the scale of the graph nevertheless the inclusion of these higher-order terms accelerates the convergence of the interpolation of the phonon spectrum with respect to the \abinitio $\qq$ grid.
As mentioned in the main article, the effect on the mobility is
estimated to be of the order of 1\%, value that is anyway larger than the one observed from the $\cal{E}$ term.
We speculate that, 
in more ``complicated" materials,
the inclusion of these higher-order terms will play a more important role for the accurate description of the phonon dispersion in the long-wavelength limit.
Figures~\ref{fig:GaP_ph_conv} and \ref{fig:GaAs_ph_conv} show the same comparison for GaP and GaAs.

\section{Convergence with respect to the DFPT mesh}

\begin{figure}
    \centering
    \includegraphics[width=0.6\textwidth]{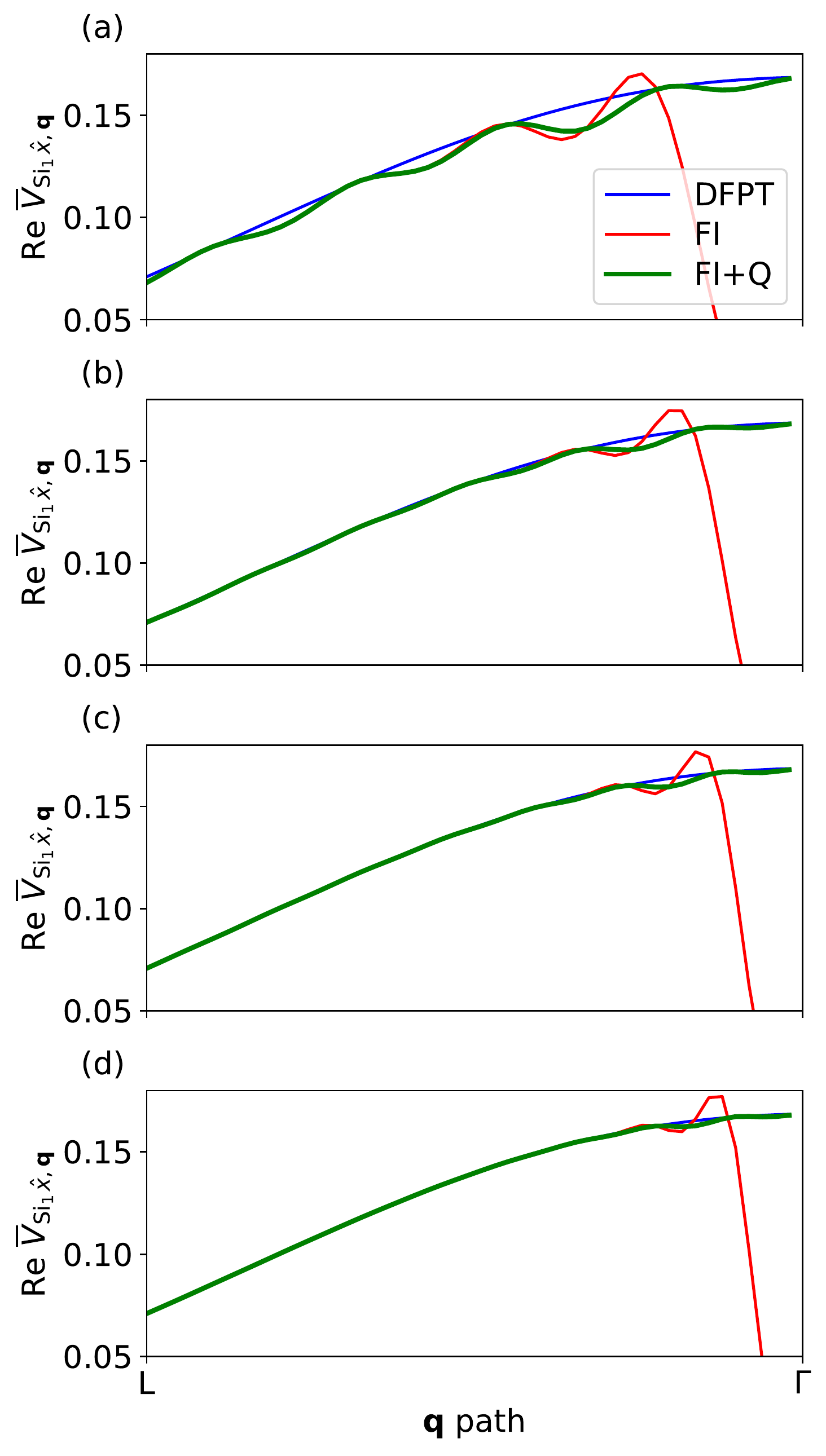}
    \caption{Close-up on the L$-\Gamma$ region of Fig.~2(c) of the main text. The average over the unit cell of the exact DFPT potentials and the Fourier interpolated result obtained with (FI+Q) and without (FI) quadrupolar contribution. The interpolation is performed starting from a (a) \grid{9}, (b) \grid{12}, (c) \grid{15} and (d) \grid{18} DFPT $\qq$ mesh.
    }
    \label{fig:Si_pot_conv}
\end{figure}

Figure~\ref{fig:Si_pot_conv} shows the interpolated potential in Si, as in Fig.~\ref{fig:Si_Si1_avg_interp}, computed from different DFPT $\qq$ meshes of increasing density. 
This shows that the wiggles observed when interpolating from a \grid{9} $\qq$ mesh disappear when a denser mesh is used. However, as shown in the following, these wiggles do not play an important role on the electron mobility.

\begin{figure}
    \centering
    \includegraphics[clip,trim=0.2cm 0.2cm 0.2cm 0.2cm,width=.7\textwidth]{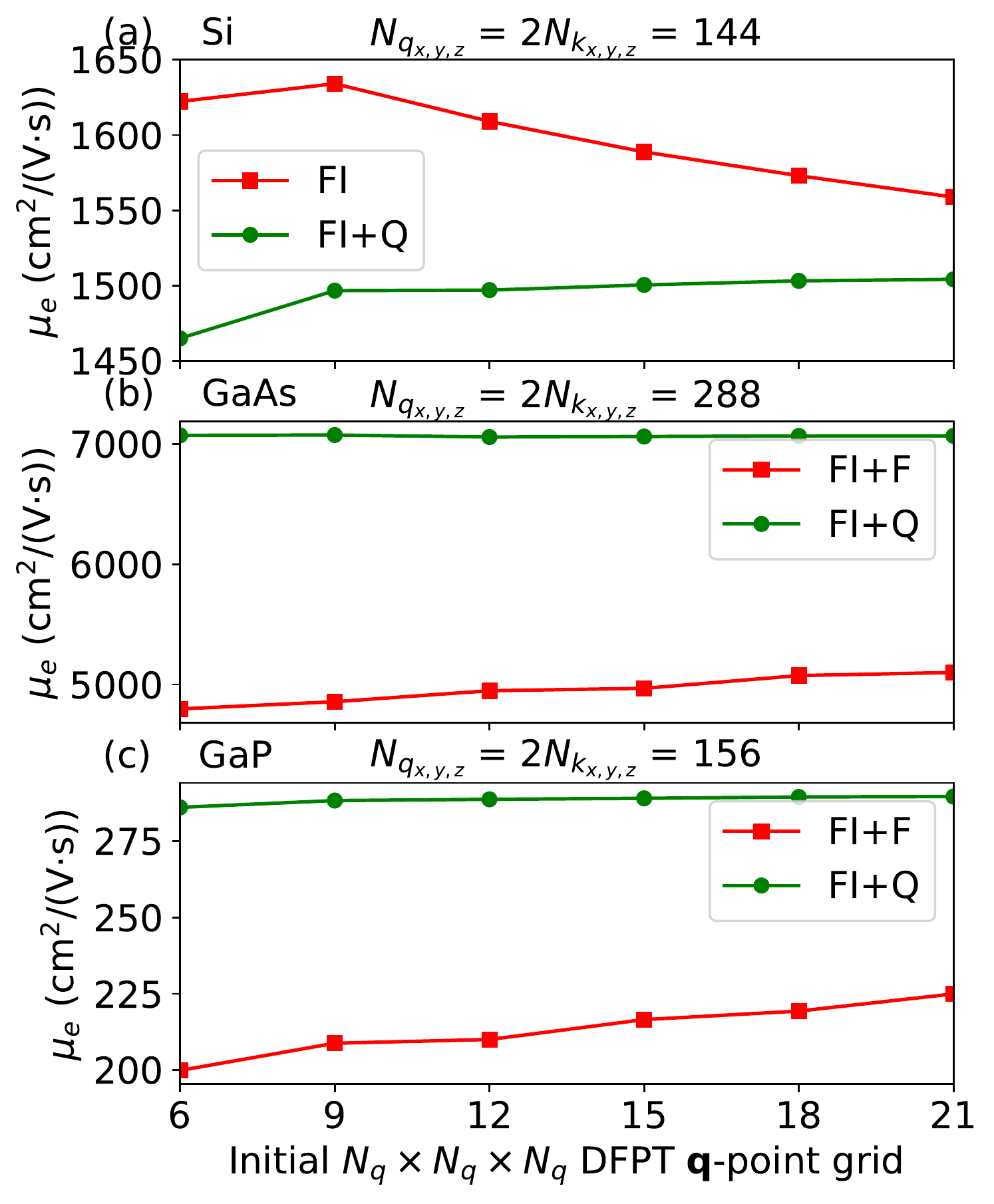}
    \caption{Convergence of the electron mobility in (a) Si, (b) GaAs and (c) GaP as a function of the initial DFPT $\qb$-point grid. The linewidths and the mobilities are obtained by interpolating from these DFPT grids to 
    (a) \grid{72} $\kb$- and \grid{144} $\qb$-point grids,
    (b) \grid{144} $\kb$- and \grid{288} $\qb$-point grids and
    (c) \grid{78} $\kb$- and \grid{156} $\qb$-point grids~\cite{Brunin2019prb}.
    The red lines (FI for Si and FI+F for GaAs and GaP) include the special treatment of the Fr\"ohlich contribution in the LR component of the potentials. The green line (FI+Q) includes the special treatment of both Fr\"ohlich and dynamical quadrupoles.}
    \label{fig:mob_dfpt_conv}
\end{figure}

\begin{figure}
    \centering
    \includegraphics[clip,trim=0.2cm 0.2cm 0.2cm 0.2cm,width=.7\textwidth]{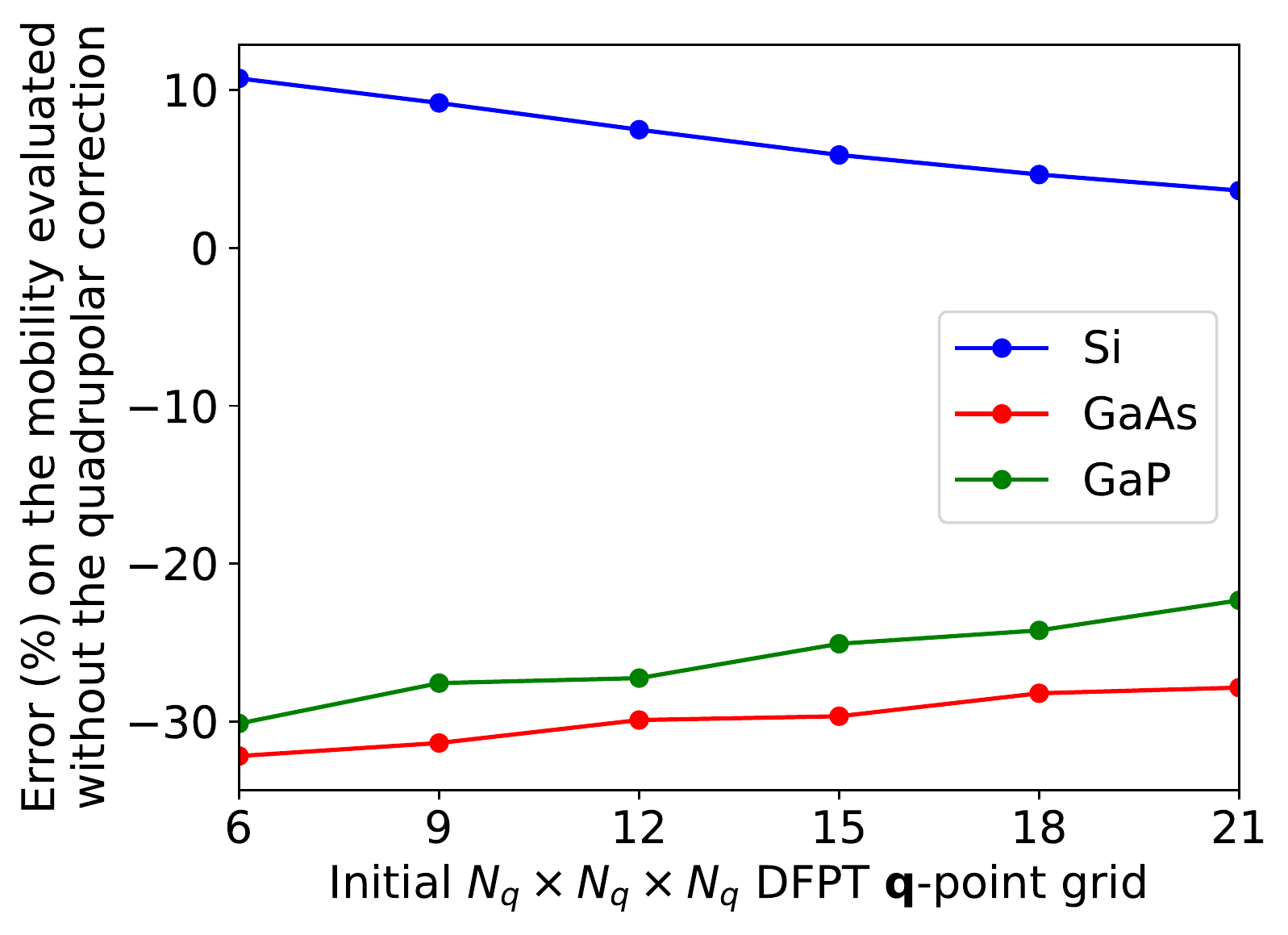}
    \caption{Error on the electron mobility in Si, GaAs and GaP evaluated without the quadrupolar correction.}
    \label{fig:error_tot}
\end{figure}

Figures~\ref{fig:mob_dfpt_conv}(a), (b) and (c) report the electron mobilities in Si, GaAs and GaP computed using plane waves with \abinit as described in Ref.~\cite{Brunin2019prb}, as a function of the initial $\qb$ grid.
For each initial \abinitio DFPT mesh, we interpolate the scattering potentials to obtain the lifetimes on 
$\kb$- and $\qb$-point meshes that are dense enough
to reach convergence in the mobility within 5\%.
Fig.~\ref{fig:error_tot} gives, for these materials and for each DFPT $\qq$ mesh, the error made when 
the quadrupole interaction is not treated correctly.
If the terms associated to the 
dynamical quadrupoles 
are included in the LR part of the potentials, 
a \grid{9} (\grid{6}) $\qb$ mesh is already
sufficiently dense in Si and GaP (GaAs) for the correct interpolation of the potentials (green lines) whereas without $\mathbf{Q}_{\kappa\alpha}$
the convergence is much slower and not even reached with a \grid{21} $\qq$ mesh (red lines).
Using $\qq$ grids typically reported in the literature (\grid{8}) can therefore lead to significant errors at the level of the mobility.

\begin{figure}
    \centering
    \includegraphics[clip=0.2cm 0.2cm 0.2cm 0.2cm,width=0.7\textwidth]{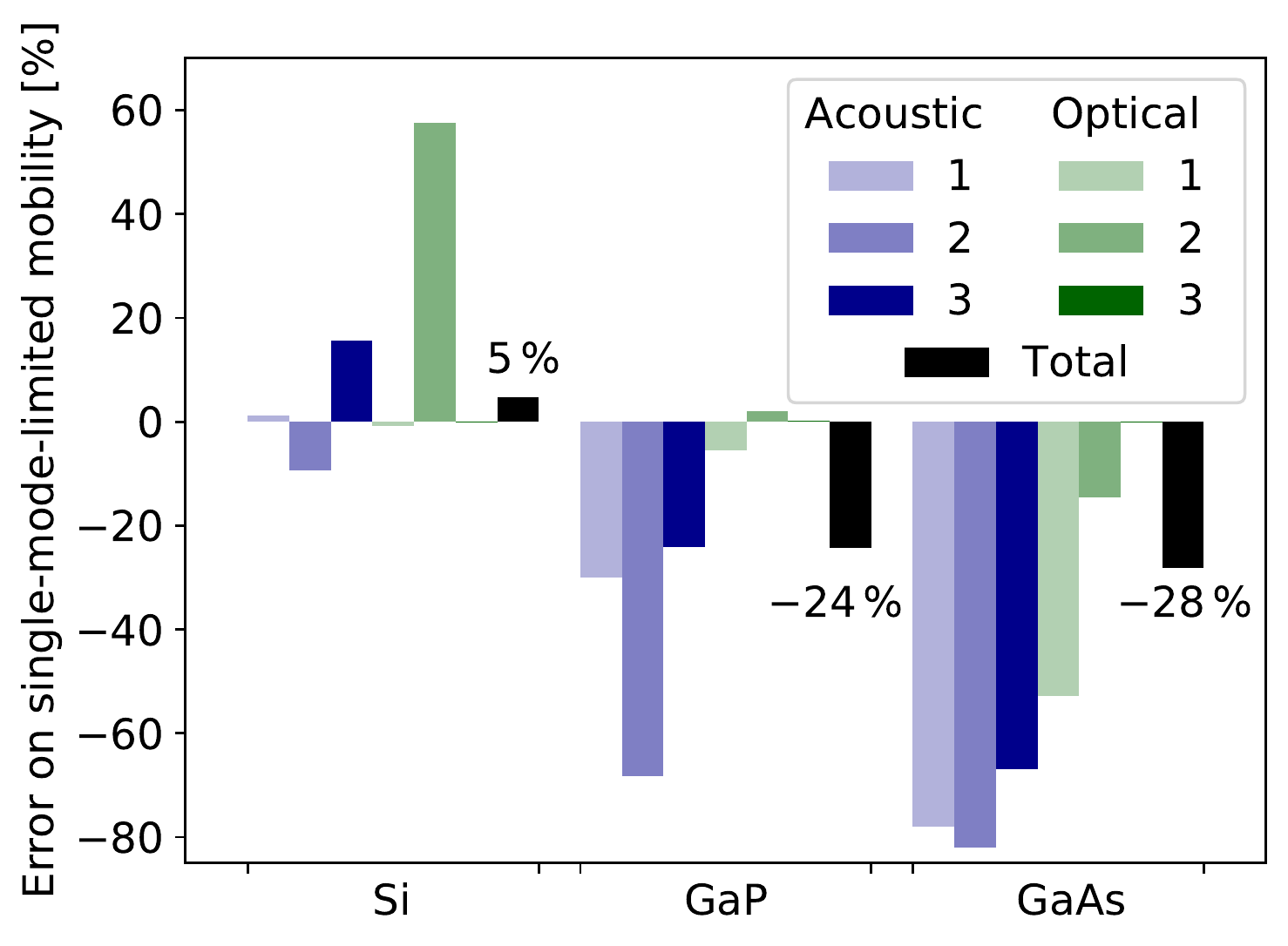}
    \caption{Error on the single-phonon-mode limited mobility when the quadrupole interaction is not correctly treated in Si, GaP and GaAs, for acoustic (blue) and optical (green) modes. The error on the total mobility is given in black. We interpolate starting from a \grid{18} DFPT mesh.}
    \label{fig:mob_modes}
\end{figure}

We have also computed the same errors for each single-mode-limited mobilities in Si, GaP and GaAs as in Fig.~3, but starting from \grid{18} DFPT $\qq$ meshes. The results are given here in Fig.~\ref{fig:mob_modes}. The errors are slightly smaller but still far from negligible, and the conclusions on the most affected modes are the same.

\section{Importance of the region around $\Gamma$}

\begin{figure}
    \centering
    \includegraphics[clip,trim=0.2cm 0.2cm 0.2cm 0.2cm,width=.6\textwidth]{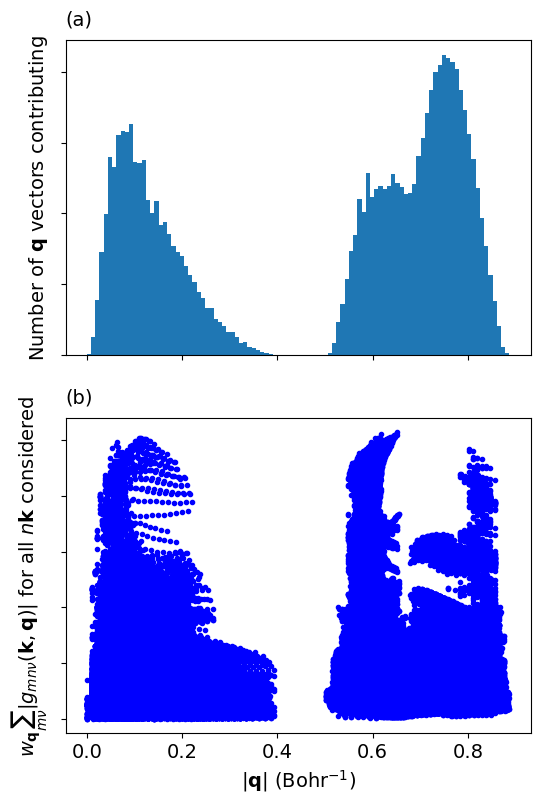}
    \caption{(a) Distribution of the number of $\qq$ points used in the integration of the electron lifetimes in Si as a function of $|\qq|$.
    (b) Values of the e-ph matrix elements $\gkkp$ summed over $m$ and $\nu$, for all the n$\kk$ states considered relevant for the computation of the mobility. The weight of the $\qq$ points are also taken into account. This figure reveals that the sampling of the region around $\Gamma$ significantly contributes to the final results.}
    \label{fig:Si_GammaRegion}
\end{figure}

\begin{figure}
    \centering
    \includegraphics[clip,trim=0.2cm 0.2cm 0.2cm 0.2cm,width=.6\textwidth]{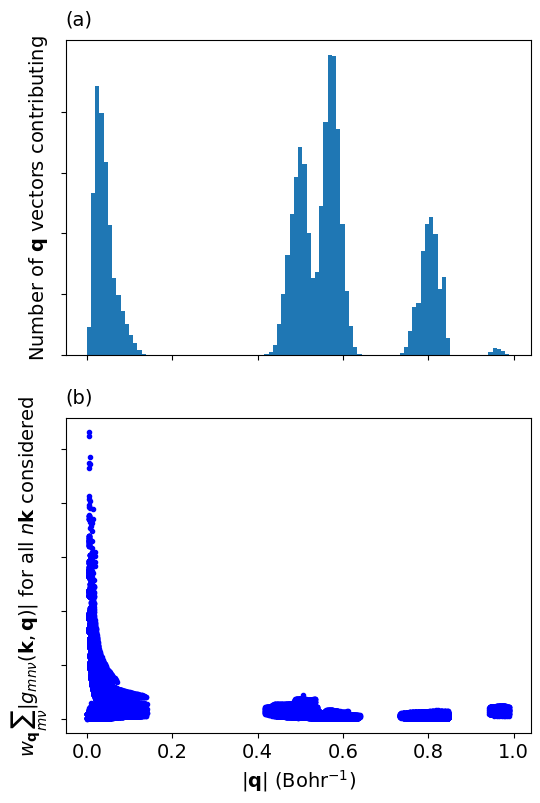}
    \caption{(a) Distribution of the number of $\qq$ points used in the integration of the electron lifetimes in GaAs as a function of $|\qq|$.
    (b) Values of the e-ph matrix elements $\gkkp$ summed over $m$ and $\nu$, for all the n$\kk$ states relevant for the computation of the mobility. The weight of the $\qq$ point is also taken into account. These show that the the sampling of the region around $\Gamma$ is very important.}
    \label{fig:GaAs_GammaRegion}
\end{figure}

Figures~\ref{fig:Si_GammaRegion} and \ref{fig:GaAs_GammaRegion} show the number of $\qq$ points included in the computation of the imaginary part of the e-ph self-energy and the sum of the absolute value of the e-ph matrix elements as a function of the length of the $\qq$ point. 
An important fraction of phonon-mediated transitions involve small momentum transfer with significant probability amplitude.
In GaAs (Fig.~\ref{fig:GaAs_GammaRegion}) 
the e-ph matrix elements show a sudden increase for small $|\qq|$
because of the Fr\"ohlich divergence.
Note that the analysis of the e-ph matrix elements takes into account crystal momentum conservation but not energy conservation in the sense that all the phonon modes $\nu$ are included in the sum of the absolute value of the e-ph matrix elements.

\bibliographystyle{apsrev4-1}